\documentclass[%
    longbibliography,
    reprint,
    nofootinbib,
    amsmath,
    amssymb,
    aps,
    physrev,
    floatfix,
    superscriptaddress
]{revtex4-2}

\usepackage{amsmath}
\usepackage{mathtools}
\usepackage{xurl}
\usepackage[title]{appendix}
\usepackage{xcolor}
\usepackage{siunitx}
\usepackage[final]{graphicx}
\usepackage{placeins}
\usepackage[utf8]{inputenc}
\usepackage{tabularray}
\usepackage[caption=false]{subfig}
\usepackage[colorlinks,citecolor=blue,linkcolor=blue,urlcolor=blue]{hyperref}
\usepackage{cleveref}
\usepackage{upgreek}
\usepackage{csquotes}

\newcommand{\I}{\mathrm{i}}
\newcommand{\E}{\mathrm{e}}
\newcommand{\dd}{\mathrm{d}}

\DeclareMathOperator{\sinc}{sinc}
\renewcommand{\vec}[1]{\boldsymbol{#1}}
\renewcommand{\Re}{\operatorname{Re}}
\renewcommand{\Im}{\operatorname{Im}}
\newcommand{\comm}[2]{[{#1},{#2}]}

\Crefname{equation}{Eq.}{Eqs.}
\Crefname{figure}{Fig.}{Figs.}
\Crefname{tabular}{Tab.}{Tabs.}
\newcommand{\crefrangeconjunction}{--}
\crefformat{pluralequation}{eqs.~(#2#1#3)}
\Crefformat{pluralequation}{Eqs.~(#2#1#3)}
\Crefrangeformat{pluralequation}{Eqs.~(#3#1#4--#5#2#6)}
\Crefmultiformat{pluralequation}{Eqs.~(#2#1#3)}{ and~(#2#1#3)}{,~(#2#1#3)}{ and~(#2#1#3)}
\Crefmultiformat{equation}{Eqs.~(#2#1#3)}{ and~(#2#1#3)}{,~(#2#1#3)}{ and~(#2#1#3)}
\Crefformat{appendix}{Appendix~#2#1#3}
\Crefformat{section}{Sec.~#2#1#3}
\Crefmultiformat{section}{Secs.~#2#1#3}{ and~#2#1#3}{,~#2#1#3}{ and~#2#1#3}

\counterwithin*{equation}{section}
\renewcommand{\theequation}{\arabic{section}.\arabic{equation}}
\let\appx\appendix
\renewcommand{\appendix}{\appx\renewcommand{\theequation}{\Alph{section}.\arabic{equation}}}

\newcommand{\of}[1]{%
    \mathchoice
        {\!\left({#1}\right)}%
        {\!\left({#1}\right)}%
        {\left({#1}\right)}%
        {\left({#1}\right)}%
}
\newcommand{\ofb}[1]{%
    \mathchoice
        {\!\left[{#1}\right]}%
        {\!\left[{#1}\right]}%
        {\left[{#1}\right]}%
        {\left[{#1}\right]}%
}

\newcommand{\tsp}[1]{%
    \mathchoice
        {{\kern#1\dimexpr0.16667em\relax}}%
        {}%
        {}%
        {}%
}
\DeclarePairedDelimiter{\abs}{{\tsp{0.25}|}}{{|\tsp{0.25}}}

\newcommand{\T}{\mathrm{T}}
\renewcommand{\H}{\mathrm{H}}

\newcommand{\Eta}{\mathrm{H}}
\newcommand{\Beta}{\mathrm{B}}

\DeclareMathOperator{\diag}{diag}

\DeclareSIUnit{\rad}{rad}

\newcommand{\occ}{\mathfrak{c}}
\newcommand{\ocg}{\mathfrak{g}}
\newcommand{\och}{\mathfrak{h}}

\renewcommand{\AA}{\mathcal{A}}
\newcommand{\OO}{\mathcal{O}}
\newcommand{\UU}{\mathcal{U}}
\newcommand{\XX}{\mathcal{X}}
\newcommand{\YY}{\mathcal{Y}}

\newcommand{\Sq}{{\mathrm{S}}}
\newcommand{\Asq}{{\mathrm{AS}}}
\newcommand{\ASq}{\Asq}

\newcommand{\Dvar}{{\Delta^2}}

\newcommand{\eqdef}{\stackrel{\mathrm{def.}}{=}}

\newcommand{\nophase}{\check}

\newcommand{\set}[1]{\{#1\}}

\ExplSyntaxOn
\NewDocumentCommand{\refcite}{m}{\refcite_aux:n{#1}}

\cs_new_protected:Npn \refcite_aux:n #1%
{%
    \seq_set_split:Nnn \l_tmpa_seq { , } { #1 }%
    \int_compare:nNnTF { \seq_count:N \l_tmpa_seq } > { 1 }%
        { Refs.\nobreakspace\cite{#1} }%
        { Ref.\nobreakspace\cite{#1} }%
}
\ExplSyntaxOff

\DeclareMathOperator{\cov}{cov}

\newcommand{\lof}{\iota}

\makeatletter
\newcommand{\Biggg}{\bBigg@{3}}
\newcommand{\vast}{\bBigg@{4}}
\newcommand{\Vast}{\bBigg@{5}}
\makeatother

\newcommand{\revA}{}

\begin{document}

  \title{Characterization of spatial Schmidt modes in high-gain SU(1,1) interferometers}

  \author{D.~Scharwald}
  \affiliation{
    Heinz Nixdorf Institute, Paderborn University, 
    Fürstenallee 11, 33102 Paderborn, Germany
  }
  \affiliation{
    Department of Electrical Engineering and Information Technology, 
    Paderborn University,
    Warburger Straße 100, 33098 Paderborn, Germany
  }
  \affiliation{
    Department of Physics, Paderborn University,
    Warburger Straße 100, 33098 Paderborn, Germany
  }
  \author{P.~R.~Sharapova}
  \affiliation{
    Department of Physics, Paderborn University,
    Warburger Straße 100, 33098 Paderborn, Germany
  }

  \begin{abstract}
    Multimode quantum light has promising applications in many areas of
    physics, such as quantum communications and quantum computing.
    However, its multimode nature also makes it challenging to measure
    its properties.
    Recently
    [\href{https://doi.org/10.1364/OPTICAQ.524682}{Optica Quantum \textbf{3}, 36 (2025)}],
    a technique for the simultaneous measurement of squeezing \revA{of multiple broadband
    modes based on a phase-sensitive 
    amplification approach was experimentally implemented using a setup that}
    effectively corresponds to an SU(1,1) interferometer.
    \revA{Here, we aim to provide a complete theoretical analysis of the modal structure of
    SU(1,1) interferometers (generally unbalanced) and a 
    detailed theoretical formal derivation of the framework} 
    for this technique. \revA{Utilizing the joint
    Schmidt decomposition of the transfer functions, we investigate the shape 
    and phase profiles of the modes of the SU(1,1) interferometer and its components 
    [parametric down-conversion (PDC) sections] for different parametric gain regimes.}
    \revA{We discover a complicated interplay between the PDC
    modes and the modes of the entire interferometer, and analyze it by
    using their overlap coefficients as a similarity measure.
    Finally, we develop a rigorous  
    processing method for the aforementioned multimode squeezing measurement technique and 
    discuss necessary approximations to make this method experimentally
    feasible.}
  \end{abstract}

  \maketitle

    \section{INTRODUCTION}%
            \label{sec:intro}\label{sec:smodespdc}

      \subsection{Motivation: Multimode squeezed light}

        In the last decade, high-gain parametric down-conversion (PDC) has
        become a topic of great interest in quantum computing, sensing and
        metrology~\cite{PhysRevA.105.043701,perez2015,PRA102,qute.202300353,%
        multimodesqueezing,PhysRevA.104.022208,PRR5}. 
        Using PDC, it is possible to prepare quantum states with large photon numbers,
        that is, macroscopic states, which still carry quantum
        correlations~\cite{PRR2,PRA91}.
        Such states also exhibit quadrature squeezing and therefore have important applications in 
        continuous variable quantum computing~\cite{RevModPhys.84.621,Larsen2021,Madsen2022}, 
        quantum metrology~\cite{Schaffrath_2024,PhysRevLett.129.121103,PhysRevApplied.16.044031}, 
        quantum cryptography~\cite{PhysRevA.61.022309},
        cluster state generation~\cite{doi:10.1126/science.aay2645,roh2024}, 
        and in the realization of Gaussian Boson Sampling~\cite{PhysRevLett.127.180502}.
        In particular, in metrology, 
        squeezed states of light make it possible to implement new
        types of interferometers, called nonlinear interferometers [SU(1,1) interferometers], 
        which consist of two consecutive PDC sections and allow
        for high sensitivity measurements which may surpass the shot 
        noise level (supersensitive
        measurements)~\cite{PhysRevA.110.022432,Manceau2017,PhysRevA.33.4033,PRR5}. 
        
        \par \emph{Multimode} squeezed
        light~\cite{Kopylov2025,PhysRevA.66.053815,multimodesqueezing,PhysRevA.73.063819,%
        PhysRevResearch.3.013199,PhysRevResearch.5.023178} 
        consists of multiple squeezed modes in a single pulse,
        each of which has its own degree of squeezing that can be controlled by proper source
        engineering~\cite{PhysRevA.97.033808,Houde2023}.
        Such states of light have important applications in quantum information 
        processing~\cite{PhysRevA.106.052607,PhysRevResearch.6.043113}
        and quantum communication~\cite{Amitonova:20} since they are very attractive for 
        scalability aspects in quantum computing and multi-parameter estimation in metrology. 
        However, the multimode nature of these states significantly complicates their description
        and characterization. 
        This is because the conventional measurement techniques for squeezing,
        such as homodyne
        detection~\cite{Eto:07,Gerry_Knight_2004,multimodesqueezing}, require 
        an increasing amount of resources as the number of modes increases. 
        
        \par The use of multimode SU(1,1) interferometers can greatly simplify 
        the problem of measuring and characterizing many modes simultaneously.
        For example, a recently developed method for performing Wigner function tomography
        based on multimode SU(1,1) interferometry has demonstrated the advantage
        of being able to simultaneously tomograph multiple modes, as well as being
        immune to detection losses~\cite{Kalash:23}.
        A similar technique was later
        used to simultaneously measure multimode squeezing of
        multimode PDC~\cite{multimodesqueezing}, where 
        the first crystal of the interferometer
        (or the first pass through the PDC section) prepares a squeezed
        vacuum state, while the second crystal (or the second pass), acts as an analyzer. 
        However, for the correct implementation of such types of interferometers,
        it is necessary to know their modal
        structure, as well as the modal structure 
        of the PDC sections that comprise them, and 
        the complete information regarding the modal 
        overlap mismatch in the system. 
      
        \par In this work, we aim to provide a thorough 
        analysis of the modal structure
        of \revA{(generally unbalanced)} multimode SU(1,1) interferometers
        \revA{in various gain-regimes,} taking into account the 
        mismatch of the squeezed light modes generated in the different PDC sections.  
        \revA{Based on this analysis, we present a rigorous and detailed derivation of
        the multimode phase-sensitive amplification technique for measuring multimode
        squeezing using direct intensity measurements.
        We start with the exact
        reconstruction of the squeezing
        as it follows naturally from the presented approach.}
        We then provide a number of simplifications 
        that allow us to significantly reduce the complexity of the presented 
        technique and make it feasible for experimental implementations.
        This simplified technique has already been successfully realized in an 
        experimental setup in \refcite{multimodesqueezing} and 
        demonstrated good agreement
        between theory and experiment.
        \revA{The discussed approach is quite general and can in principle be applied}
        to arbitrary multimode 
        squeezed states of light and various interferometric configurations where 
        the mode-matching problem is a key bottleneck.
        \revA{In particular, this approach could be useful for systems
        where a full measurement of the states carrying squeezing is not 
        accessible to standard homodyne 
        detection~\cite{PhysRevResearch.3.013199, PhysRevResearch.5.023178}.}

        \par \revA{In order to
        provide a self-contained description of generally 
        unbalanced SU(1,1) interferometers, in the following
        section, we will briefly recapitulate important concepts for the
        general description of multimode SU(1,1) interferometers.
        The mentioned equations will be later directly used and expanded 
        upon in the following sections.}

      \subsection{Theoretical framework and Schmidt modes}

        \par In order to describe the parametric down-conversion (PDC) process 
        in both the low-gain and high-gain regime,
        we utilize a description of PDC using
        a system of coupled integro-differential equations 
        which describe
        the propagation of the plane-wave operators~$\hat{a}_s$ 
        and~$\hat{a}_i$ through the PDC section,
        where the labels~$s$ and~$i$ refer to the signal and idler beam, respectively.
        The derivation can be found in \refcite{PRR2}.
        For simplicity, in this work, we will restrict ourselves to the
        description of frequency-degenerate
        \mbox{type-I} PDC, meaning that the signal and idler photons are
        indistinguishable
        and the signal and idler labels are redundant. However, to 
        be consistent with the literature,
        we will keep the $s$ and $i$ labels for now.

        \revA{The solution of this integro-differential equation 
        approach takes the form of the well-known input-output
        relations~\cite{PRR2,PRR5}:
        \begin{subequations}
            \begin{align}
                \begin{split}
                    \hat{a}_s^{\left(\mathrm{out}\right)}\of{q_s}
                        =&\int\!\dd q_s'\,\tilde{\eta}\of{q_s,q_s'}
                        \hat{a}_s^{\left(\mathrm{in}\right)}\of{q_s'} \\
                        &+ \int\!\dd q_i'\,\beta\of{q_s,q_i'}
                        \left[\hat{a}_i^{\left(\mathrm{in}\right)}\of{q_i'}\right]^{\dagger},
                    \label{eq:as_dagger_sol}
                \end{split}\\
                \begin{split}
                    \left[\hat{a}_i^{\left(\mathrm{out}\right)}\of{q_i}\right]^{\dagger} 
                        =&\int\!\dd q_i'\,
                        \tilde{\eta}^{*}\of{q_i,q_i'}
                        \left[\hat{a}_i^{\left(\mathrm{in}\right)}\of{q_i'}\right]^{\dagger} \\
                        &+ \int\!\dd q_s'\,\beta^{*}\of{q_i,q_s'}
                        \hat{a}_s^{\left(\mathrm{in}\right)}\of{q_s'},
                        \label{eq:ai_dagger_sol}
                \end{split}
            \end{align}
        \end{subequations}
        where the variables $q_{s/i}$ label the transverse wave-vector components
        of the plane-wave modes associated with the $\hat{a}_{s/i}$ operators and
        where the plane-wave operators marked with the superscript~\textsuperscript{(in)}
        are the input operators of the PDC section which form the boundary
        condition for the set of integro-differential equations,
        while the operators marked with~\textsuperscript{(out)} are the output
        operators corresponding to the solutions of the set of integro-differential
        equations. These two types of operators are connected via the
        complex-valued transfer functions $\beta$ and $\tilde{\eta}$.
        which describe how the input plane-wave modes
        are connected to the output plane-wave modes of the PDC section
        and contain the full information regarding the PDC interaction.
        Note that this form of the solution already assumes that the signal and
        idler photons are indistinguishable~\cite{PRR5,diss}.}

        \par \revA{For a  Gaussian pump of the form
        $E_p\of{x,z,t}=E_0\E^{-\frac{x^2}{2\sigma^2}}\E^{\I\left(k_p z-\omega_p t\right)}$,
        the transfer functions can be found as the solution of
        the integro-differential equations~\cite{PRR5}
        \begin{subequations}
            \label{eq:integ_diffeqs_eta_beta_both}
            \begin{align}
                \begin{split}
                    \frac{\dd \beta\of{q_s,q_i',L}}{\dd L} &= \Gamma \int\!\dd q_i\,
                        \E^{-\frac{\left(q_s+q_i\right)^2\sigma^2}{2}} \\
                    &\qquad\times h\of{q_s,q_i,L} \tilde{\eta}^{*}\of{q_i,q_i',L},
                    \label{eq:integ_diffeqs_eb_beta}
                \end{split}\\
                \begin{split}
                    \frac{\dd \tilde{\eta}^{*}\of{q_i,q_i',L}}{\dd L} &= \Gamma \int\!\dd q_s\,
                        \E^{-\frac{\left(q_s+q_i\right)^2\sigma^2}{2}} \\
                    &\qquad\times h^{*}\of{q_s,q_i,L} \beta\of{q_s,q_i',L}.
                    \label{eq:integ_diffeqs_eb_eta}
                \end{split}
            \end{align}
        \end{subequations}
        These equations can be solved using standard methods such as the
        Runge-Kutta methods for solving ordinary differential equations
        or more elaborate solvers (see for example \Cref{sec:beta_asym}).
        Note that generally, the equations include a two-dimensional transverse 
        wave vector~$\vec{q}$.
        However, below we will consider only systems with the azimuthal 
        symmetry and, therefore, reduce our consideration to a single 
        transverse component of the wave vector.}
        The parameter $\Gamma$ is connected to the parametric
        gain of the PDC interaction, which will be elaborated on in
        \Cref{sec:single_crystal}.
        Note that the full width at half maximum (FWHM) of the pump intensity distribution
        is given by $2\sigma\sqrt{\ln 2}$.
        The function~$h$ describes the phase-matching
        inside the PDC section and is symmetric due to 
        the fact that we consider frequency-degenerate 
        \mbox{type-I} PDC; concrete expressions for $h$ will be 
        given and explained below in 
        \Cref{sec:single_crystal,sec:interferometers}.

        \par It is well known that the transfer functions for this system
        admit a \emph{joint Schmidt decomposition} \revA{(Bloch-Messiah reduction)} of the 
        form~\cite{PRR5,Christ_2013,PRA102}:
        \begin{subequations}
            \begin{align}
                \beta\of{q,q'} &= \sum_{n} \sqrt{\Lambda_{n}} u_{n}\of{q}\psi_{n}\of{q'},
                    \label{eq:jointsdecompbeta} \\
                \tilde{\eta}\of{q,q'} &= \sum_{n} \sqrt{\tilde{\Lambda}_{n}}
                    u_{n}\of{q}\psi_{n}^{*}\of{q'},
                    \label{eq:jointsdecompeta}
            \end{align}
        \end{subequations}
        where the $u_n\of{q}$ modes are the output Schmidt modes
        (connected to the output Schmidt operators, see below) and the
        $\psi_n\of{q'}$ are the input Schmidt modes (connected to
        the input Schmidt operators), while
        $\Lambda_n$ and $\tilde{\Lambda}_n$
        are the Schmidt eigenvalues associated with the two transfer functions
        $\beta$ and $\tilde{\eta}$, respectively.
        Note that the eigenvalues are connected via~\cite{Christ_2013,PRR5}:
        \begin{align}\label{eq:connection_L_Ltilde}
            \tilde{\Lambda}_n = \Lambda_n+1.
        \end{align}
        More details on the joint Schmidt decomposition
        can be found in \Cref{sec:jsdecomp}.
        \par By plugging both \Cref{eq:jointsdecompbeta,eq:jointsdecompeta} into
        the form of the solutions of the plane-wave operators,
        \Cref{eq:as_dagger_sol,eq:ai_dagger_sol}, multiplying both
        sides with $u_l^{*}\of{q_s}$ and $u_l\of{q_i}$, respectively,
        and integrating over $q_s$ and $q_i$, respectively,
        the well-known Bogoliubov transformation for the Schmidt
        operators is recovered~\cite{PRA91,Christ_2013}:
        \begin{align}\label{eq:bogoliubov_plain}
            \hat{A}_l^{\left(\mathrm{out}\right)} &= 
            \sqrt{\tilde{\Lambda}_l}\,\hat{A}_l^{\left(\mathrm{in}\right)} + \sqrt{\Lambda_l}\left[\hat{A}_l^{\left(\mathrm{in}\right)}\right]^{\dagger},
        \end{align}
        and analogously for the creation operator,
        where the input and output Schmidt operators
        are connected to
        the mode functions $\psi$ and $u$, 
        respectively~\cite{Christ_2013}:
        \begin{subequations}
            \begin{align}
                \hat{A}_l^{\left(\mathrm{in}\right)} &= \int\!\dd q\,\psi_l^{*}\of{q}
                    a^{\left(\mathrm{in}\right)}\of{q}, \\
                \hat{A}_l^{\left(\mathrm{out}\right)} &= \int\!\dd q\,u_l^{*}\of{q}
                    a^{\left(\mathrm{out}\right)}\of{q}.
            \end{align}
        \end{subequations} 
        Note that as mentioned above, the signal and idler photons are 
        indistinguishable, meaning that the transformation of
        the Schmidt operators is fully described by \Cref{eq:bogoliubov_plain}
        and its hermitian conjugate.
        The Bogoliubov transformation for the Schmidt
        operators provides a diagonalization of the input-output
        relations, thus coupling
        input and output Schmidt modes with only the same 
        mode index $l$. 

    \section{SINGLE CRYSTAL}\label{sec:single_crystal}
        \revA{Before considering the modal structure of the full SU(1,1) interferometer,}
        in this section, 
        we analyze the modal structure of light
        generated \revA{in its first half, namely} 
        a single PDC section, at low and high parametric gain. 
        For a single crystal, the function $h$ describing the phase matching and
        appearing in \Cref{eq:integ_diffeqs_eb_beta,eq:integ_diffeqs_eb_eta}
        is given by~\cite{PRR2,PRR5}:
        \begin{align}
            h\of{q_s,q_i,L} &= \E^{\I \Delta k\of{q_s,q_i} L},
        \end{align}
        where $\Delta k\of{q_s,q_i}= \sqrt{k_p^2-\left(q_s+q_i\right)^2}- \sqrt{k_s^2-q_s^2} - \sqrt{k_i^2-q_i^2}$
        is the collinear wavevector mismatch inside the crystal.
        The integro-differential equations are numerically integrated
        over the interval $\left[0,L_1\right]$ for different values
        of the parametric gain.
        The connection between the theoretical parameter $\Gamma$ and
        the experimental gain is obtained analogously to the
        experimental procedure, namely by performing a fit of
        the collinear
        output intensity $\langle\hat{N}_s\of{q=0}\rangle\,\dd q$
        with a function of the form $y\of{\Gamma}=B\sinh^2\of{A\Gamma}$, 
        where $A$ and $B$ are the fitting parameters. The experimental gain is then defined as
        $G_{\mathrm{exp}}=A\Gamma$~\cite{PRR2,PRR5,PRA91}. 
        After obtaining the transfer functions by solving
        \Cref{eq:integ_diffeqs_eb_beta,eq:integ_diffeqs_eb_eta}, 
        we can compute the joint
        Schmidt decomposition as written in
        \Cref{eq:jointsdecompbeta,eq:jointsdecompeta} and 
        \revA{obtain the Schmidt mode functions
        $u_n\of{q}$ and $\psi_n\of{q'}$. In order to provide 
        an analysis of the modal similarity of the input and
        output modes, we define the
        \emph{same-crystal overlap coefficient} $\occ_{lm}$ between these
        two modes at different indices as follows:
        \begin{align}\label{eq:samecrystaloverlapc}
            \occ_{lm} &\eqdef \int\!\dd q\, u_l\of{q} \psi_m^*\of{q}.
        \end{align}}

        \par \revA{For the
        numerical simulations, we consider a BBO crystal with a length of
        $L_1=\SI{3}{\milli\metre}$. 
        The half width of the waist ($1/\E^2$-radius) of the 
        intensity distribution of the spatial Gaussian pump envelope is chosen as 
        $\SI{70}{\micro\metre}$, meaning that
        $\sigma=\SI[parse-numbers=false]{(70/\sqrt{2})}{\micro\metre}$ in
        \Cref{eq:integ_diffeqs_eb_beta,eq:integ_diffeqs_eb_eta}, 
        which corresponds to a FWHM of the intensity distribution
        of $\SI[parse-numbers=false]{70\sqrt{2\ln2}}{\micro\metre}
        \approx\SI{82.4}{\micro\metre}$ (similar to \refcite{multimodesqueezing}). 
        For these parameters, we found the gain fitting constant as 
        $A=142.12$. Unlike in \refcite{PRR5}, where different fitting constants
        were used depending on the experimental gain regime, in this work,
        we take the same fitting constant
        for both low and high gain. This will not affect
        the following results qualitatively,
        but simplifies the numerical calculations
        greatly.}
        
        \begin{figure*}[hp!]%
            \centering%
            \includegraphics[width=0.95\linewidth]%
                {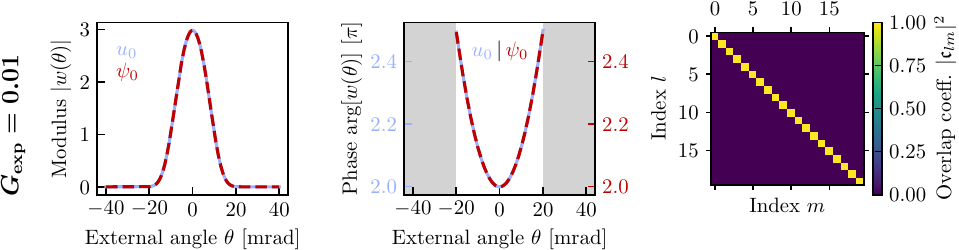}\\%
            \includegraphics[width=0.95\linewidth]%
                {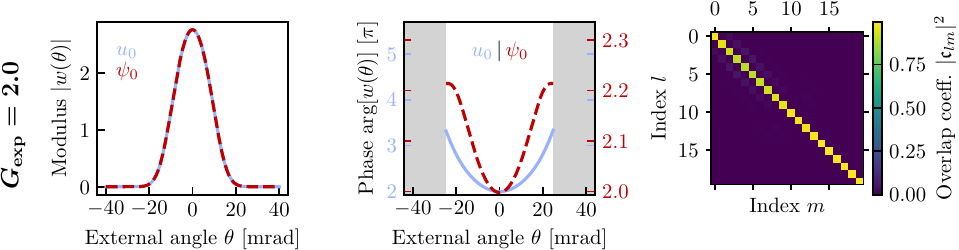}\\%
            \includegraphics[width=0.95\linewidth]%
                {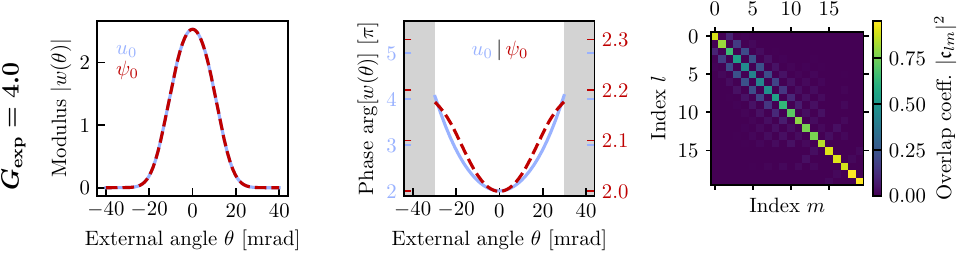}\\%
            \includegraphics[width=0.95\linewidth]%
                {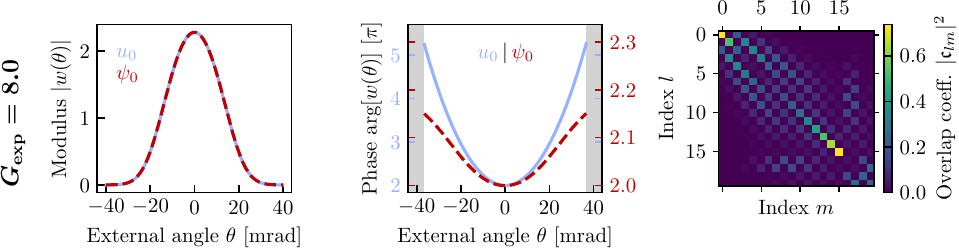}%
            \caption{\label{fig:modesandphasesc1}%
                Modulus (left column) and phase distributions
                (center column) of the first input ($w=\psi_0$, red dashed curve) and
                output ($w=u_0$, blue solid curve) Schmidt mode at different experimental gain
                values as written on the left.
                The gray margins indicate the range after which the modulus of the
                modes has decreased by a factor of $10^{-3}$ compared to the peak value.
                Since the modulus decreases,
                inside the margins, the phase remains undefined and
                is characterized by random $\pi$ (due to the fact that
                $w\approx 0$) or $2\pi$ (due to the periodicity) phase
                jumps (not shown).
                As the gain increases,
                the width of the modes increases, as already described in
                \refcite{PRR2}. At low gain, $G_{\mathrm{exp}}=0.01$,
                both input and output modes fully coincide, while at high gain, only the
                modulus of the modes coincides approximately. 
                Starting at $G_{\mathrm{exp}}=2.0$, the phase distribution
                of the input and output modes differs significantly.
                Right column: Modulus squared $\abs{\occ_{lm}}^2$ 
                of the overlap
                coefficient $\occ_{lm}$ as defined in
                \Cref{eq:samecrystaloverlapc}.
                Clearly, for low gain, the modes
                are identical,
                as indicated by the fact
                that $\abs{\occ_{lm}}^2$ is diagonal, and become
                increasingly different as the gain increases.
                Note that
                similar observations regarding the 
                similarity of the modulus of the input and output modes
                and their broadening have been made in \refcite{Christ_2013}
                for the frequency domain.
            }%
        \end{figure*}%
        
        Plots of the moduli and the phase distributions of the 
        first input ($\psi_0$) and output ($u_0$) mode
        are shown in \Cref{fig:modesandphasesc1}.
        Both functions are plotted over the
        external angle $\theta\approx q/k^{\mathrm{vac}}$
        where $k^{\mathrm{vac}}$ is the 
        modulus of the wave vector of the signal (idler)
        photons in vacuum.
        One can notice that for low parametric gain,
        $G_{\mathrm{exp}}=0.01$, the first input and
        output mode ($n=0$) coincide. 
        Additionally, from the plot of the
        $\occ_{lm}$ coefficients, it is
        clear that in the low-gain regime
        this also holds for higher-order modes.
        This can also be seen analytically from the
        Schmidt-mode theory presented in 
        \refcite{PRA91}: For low parametric gain,
        the two-photon amplitude (TPA)
        $F\of{q_s,q_i}$
        and the transfer
        function $\beta\of{q_s,q_i}$
        coincide~\cite{APL2025,diss}. If the 
        signal and idler photons are 
        indistinguishable, the TPA is symmetric in
        its arguments. This means that
        $\beta\of{q_s,q_i}$ is also symmetric in its
        arguments and admits a decomposition
        of the form\footnote{\revA{Numerically, the decomposition
        of $\beta$ as written in \Cref{eq:jointsdecompbeta_lgsymm}
        corresponds to the Takagi/Autonne decomposition~\cite{Houde2024}.}}%
        \begin{align}\label{eq:jointsdecompbeta_lgsymm}
            \beta\of{q,q'} &= \sum_{n}
                \sqrt{\Lambda_{n}} u_{n}\of{q} u_{n}\of{q'}.
        \end{align}
        However, the decomposition for $\beta$ as
        written in \Cref{eq:jointsdecompbeta_lgsymm}
        is, on its own, not sufficient to conclude
        that the input and output modes of the system 
        are identical.
        To conclude this, the full joint decomposition
        including $\tilde{\eta}$ must be investigated.
        However, clearly, for low gain,
        $\tilde{\eta}\of{q,q'}\approx\delta\of{q-q'}$
        and $\sqrt{\tilde{\Lambda}_{n}}\approx 1$,
        meaning that $\tilde{\eta}$ may be decomposed
        as 
        \begin{align}
            \tilde{\eta}\of{q,q'} &\approx \sum_{n}
            u_{n}\of{q} u_{n}^{*}\of{q'},
                \label{eq:jointsdecompeta_lgsymm}
        \end{align}
        for \emph{any} orthonormal basis
        with functions $u_{n}$,
        confirming the validity of \Cref{eq:jointsdecompbeta_lgsymm}.

        \par As the parametric gain increases,
        the mode profiles (moduli of the modes)
        broaden, which has already been observed
        in \refcite{PRR2}. Moreover, for any gain,
        the moduli of the first input ($\abs{\psi_0}$)
        and output mode ($\abs{u_0}$)
        coincide to a good approximation, which 
        has also been demonstrated for example in 
        \refcite{Christ_2013} in the
        frequency domain. 
        \revA{However, the phase profiles of the Schmidt modes 
        in the high-gain regime have not previously been studied in 
        detail and are here presented in \Cref{fig:modesandphasesc1}.
        Clearly, the
        phase profiles of the input and output modes} differ significantly with
        increasing gain and the
        input modes have a much flatter phase
        profile than the output modes.

        \par For the high-gain regime, the difference 
        in the phases between the input and output modes 
        seems to be a manifestation of 
        time ordering effects: The photons generated at
        the beginning of the crystal
        are distinguished by their phase from those 
        generated at the end.
        These effects become especially relevant at
        high parametric gain, where the plots of
        the overlap coefficient $\occ_{lm}$ explicitly 
        demonstrate that the
        input and output modes are no longer identical. 
        However, as the mode index increases, the number of 
        photons in the mode decreases. 
        \revA{As such, it may be expected that for large mode indices,
        the same-crystal overlap coefficient again approaches a diagonal
        form (as in the low-gain regime). 
        Indeed, this behavior can be observed in \Cref{fig:modesandphasesc1}: 
        For $G_{\mathrm{exp}}=4$, $\abs{\occ_{lm}}^2$ approaches a diagonal form
        after around~$l=m=6$, while for $G_{\mathrm{exp}}=8$, it can be seen
        that the overlap coefficients becomes more diagonal approaching $l=m=15$.
        The behavior of the modes $n>15$ is investigated in more detail below.}

        \par
\begin{figure*}[!ht]%
            \centering%
            \includegraphics[width=\linewidth]{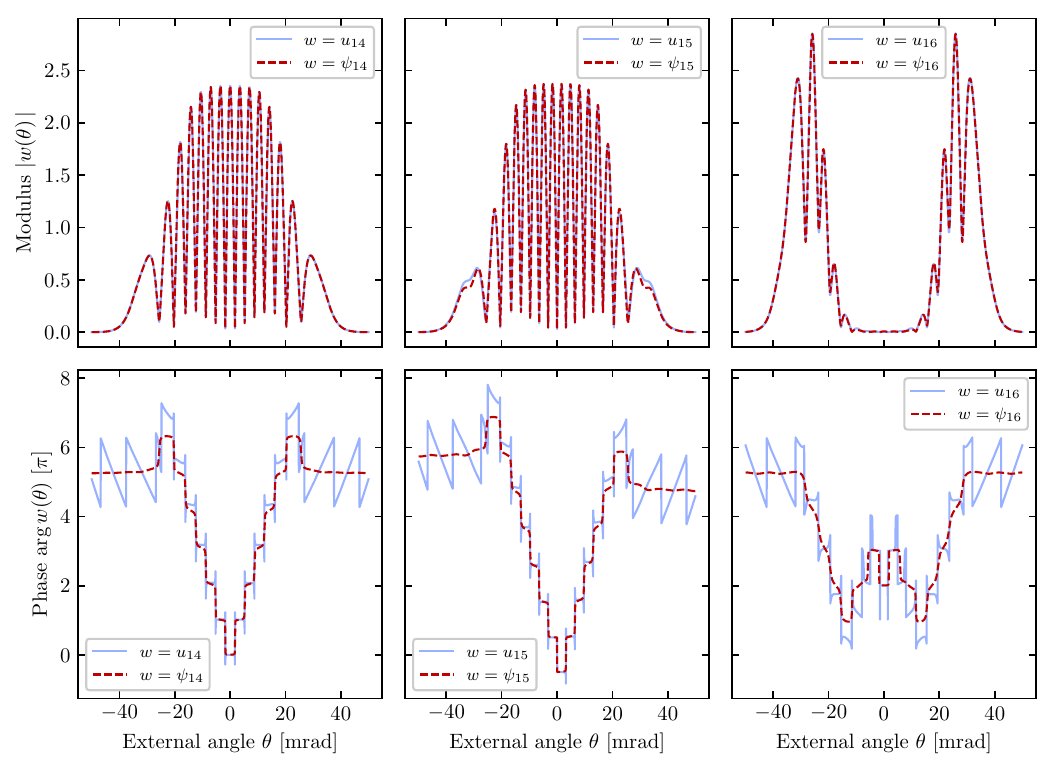}
            \caption{\label{fig:moremodes}%
                First row: Plots of the modulus of several higher-order input and
                output modes of a 
                single BBO crystal at high gain $G_{\mathrm{exp}}=8$. 
                Second row: Phase distributions of the corresponding mode functions.
                Evidently, the moduli
                of the input and output modes $\psi_{15}$ and $u_{15}$ do 
                not coincide
                at around $\theta=\pm\SI{35}{\milli\rad}$. Between modes
                $n=15$ and $n=16$,
                a structural change in the mode shapes occurs: For 
                $n\leq 15$, the moduli of the
                modes are similar to Hermite-Gaussian modes and are 
                mostly localized near the collinear direction ($\theta=0$),
                while for $n=16$, the modulus of the modes vanishes 
                near the collinear
                direction and is mostly nonzero between $-40$ and 
                $\SI{-20}{\milli\rad}$
                and from $20$ to $\SI{40}{\milli\rad}$. See
                also \Cref{fig:modesandphasesc1}.
            }%
        \end{figure*}

 It should be noted that generally, at high gain, both the modulus and
        the phase profile of the input and output modes are different,
        even though they seem to be identical for lower-order modes.
        This can be seen explicitly from
        \Cref{fig:moremodes}, which shows the moduli of the input and output
        modes with indices $14$, $15$ and $16$ at high
        gain, $G_{\mathrm{exp}}=8$. 
        Clearly, for the
        input and output modes $\psi_{15}$ and $u_{15}$, the modulus does not
        coincide for broad angles around $\theta=\pm\SI{35}{\milli\rad}$. 
        The differences in the input and output modes
        can be explained by inspecting
        the transfer function $\beta$ in mode detail, see \Cref{sec:beta_asym}.

        \par Additionally, \Cref{fig:moremodes,fig:modesandphasesc1} 
        demonstrate that the modes with $n\leq15$ are mostly localized
        near the collinear direction and have profiles similar to the 
        Hermite-Gaussian modes: Their number of peaks increases 
        as the mode index is increased. However, 
        this is no longer the case for $n=16$, since the 
        PDC intensity distribution has a $\sinc$-shape~\cite{PRR2,PRR5}.
        This change in the general shape of the mode explains the 
        behavior of the overlap
        matrix $\occ_{lm}$ after $l=m=15$, see \Cref{fig:modesandphasesc1}.
        Clearly, since the mode with $m=16$ is no longer localized near the 
        collinear direction, it may share overlap with several low-order modes,
        which all have a small modulus near these broader angles around
        $-40$ to $\SI{-20}{\milli\rad}$ and $20$ to $\SI{40}{\milli\rad}$.
        At the same time, there is a strong decrease in the
        overlap between the input and output modes at $l=m=16$, which
        is related to the phase profile of the modes~\cite{diss}.
        Indeed, as shown in \Cref{fig:moremodes}, the relative phase 
        between the modes is almost flat for $l=m=15$ in the region 
        where the intensity of the modes is nonzero, which is not 
        the case for $l=m=16$. Here, instead, the relative phase has 
        an approximately linear dependence on the angle for
        the region where the intensity distribution
        is nonzero, which causes the overlap integral between 
        the modes to drop to almost zero. 
        \revA{Note that the index of the mode at which the flip 
        from the Hermite-Gaussian profiles to the double-peak profiles
        occurs depends on the chosen system parameters.}

    \section{MODAL STRUCTURE OF SU(1,1) INTERFEROMETERS}\label{sec:interferometers}

\begin{figure}%
            \centering%
            \begingroup%
            \def\svgwidth{3.4in}%
            \fontsize{7pt}{7pt}\selectfont%
            \input{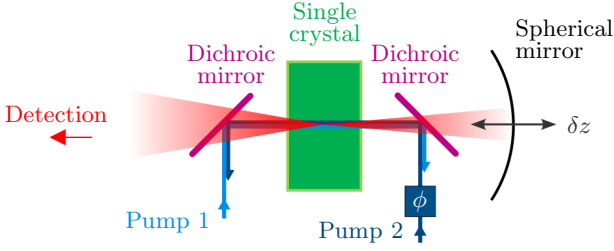}%
            \endgroup%
            \caption{\label{fig:su11setup}
                Sketch of the SU(1,1) interferometer
                setup in \emph{reflection geometry} as considered
                in this work. A single crystal is pumped with
                Pump~1, corresponding to a parametric gain $G^{\left(1\right)}$ (first pass).
                The generated PDC radiation is then reflected back onto the crystal by
                a spherical mirror. Pump~1 is removed from the setup and Pump~2 is 
                fed into the crystal using two dichroic mirrors, so that the 
                parametric gain $G^{\left(2\right)}$ of the second pass may differ
                from the first pass. Additionally, the interferometric phase $\phi$
                is varied by manipulating the phase of Pump~2.
                If $G^{\left(1\right)}\neq G^{\left(2\right)}$, the spherical mirror may
                be shifted a distance $\delta z$ towards or away from the crystal
                in order to improve the interferometric visibility,
                see the discussion in \Cref{sec:unbalanced_interf} and
                \Cref{fig:visoverdeltaz}.
                Note that the diffraction of the PDC radiation is shown exaggerated
                with respect to the rest of the figure elements.
                See also \refcite{PRR5,multimodesqueezing} for similar figures.
            }
        \end{figure}

        In the following, we turn to SU(1,1) interferometers consisting of two PDC sections.
        We will use the superscript~\textsuperscript{(1)} for quantities referring
        to the first crystal and, analogously,~\textsuperscript{(2)} for quantities
        related to the second crystal and~\textsuperscript{(SU)} for quantities describing
        the interferometer as a whole. A general setup of an SU(1,1) interferometer is
        shown in \Cref{fig:su11setup}. Unless specified otherwise, the parameters
        will be the same as for the single crystal case discussed in \Cref{sec:single_crystal}.

        \subsection{Balanced fully compensated interferometer}%
                \label{sec:perfect_compens_interf}
            \revA{In this section, we revisit the properties of the compensated SU(1,1) 
            interferometer based on \refcite{PRR5} and extend its description by accounting 
            for the modal structure of the interferometer 
            and its two PDC sections.}
            For a balanced, lossless and fully 
            compensated SU(1,1)
            interferometer, the functions $h$ describing the
            phase matching are given by~\cite{PRR5}:
            \begin{subequations}
                \begin{align}
                    h^{\left(1\right)}\of{q_s,q_i,L}
                        &= \E^{\I\Delta k L}, \label{eq:pm_func_c1_ideal} \\
                    h^{\left(2\right)}\of{q_s,q_i,L}
                        &= \E^{-\I \Delta k\of{q_s,q_i}
                        \left[L-2L_1\right]} \E^{\I \phi},
                        \label{eq:h2_comp}
                \end{align}
            \end{subequations}
            for the first and second crystal, respectively, 
            where $\phi$ is the relative phase difference introduced
            to the pump, signal and idler radiation.
            Due to the perfect
            diffraction compensation, this phase $\phi$ does
            not depend on the transverse wave vectors.
            It should be noted that
            in \Cref{eq:pm_func_c1_ideal,eq:h2_comp},
            the phase matching functions are written in a
            form which implies that the integration domain
            is given by the interval $[0,L_1]$
            for both crystals, where $L_1$ is the crystal length
            of each crystal.

            \par As was already described
            in \refcite{PRR5}, the focusing
            element used for the diffraction compensation
            introduces a symmetry in the system, which means that the
            entire interferometer is fully described by the properties of the
            \revA{first} crystal and the contribution of the flat phase $\phi$, which
            does not depend on the transverse wave vectors.
            By expressing the transfer functions of the second crystal
            in terms of those of the first crystal~\cite{PRR5}
            \begin{subequations}
                \begin{align}
                    \beta^{\left(2\right)}\of{q,q'} &=
                        \E^{\I\phi}\beta^{\left(1\right)}\of{q',q}, \label{eq:conn_tf1tf2_beta} \\
                    \tilde{\eta}^{\left(2\right)}\of{q,q'} &=
                        \left[\tilde{\eta}^{\left(1\right)}\of{q',q}\right]^{*}, \label{eq:conn_tf1tf2_eta}
                \end{align}
            \end{subequations}
            and comparing these expressions with the
            joint Schmidt decomposition for the first crystal 
            as written in
            \Cref{eq:jointsdecompbeta,eq:jointsdecompeta},
            it is immediately obvious 
            that the joint decomposition for the
            transfer functions of
            the second crystal is given by
            \begin{subequations}
                \begin{align}
                    \beta^{\left(2\right)}\of{q,q'} &= \sum_{n} \sqrt{\Lambda_{n}^{\left(1\right)}}
                        \left[\E^{\frac{\I}{2}\phi} \psi_{n}^{\left(1\right)}\of{q}\right]
                        \left[\E^{\frac{\I}{2}\phi} u_{n}^{\left(1\right)}\of{q'}\right],
                        \label{eq:sdecomp_beta} \\
                    \tilde{\eta}^{\left(2\right)}\of{q,q'} &= \sum_{n} \sqrt{\tilde{\Lambda}_{n}^{\left(1\right)}}
                        \left[\E^{\frac{\I}{2}\phi} \psi_{n}^{\left(1\right)}\of{q}\right]
                        \left[\E^{\frac{\I}{2}\phi} u_{n}^{\left(1\right)}\of{q'}\right]^{*}.
                        \label{eq:sdecomp_eta}
                \end{align}
            \end{subequations}
            Furthermore, it follows that the eigenvalues of both crystals
            are connected via:
            \begin{subequations}
                \begin{align}
                    \Lambda_{n}^{\left(2\right)} &= \Lambda_{n}^{\left(1\right)}, \\
                    \tilde{\Lambda}_{n}^{\left(2\right)} &= \tilde{\Lambda}_{n}^{\left(1\right)}, \\
            \intertext{while the Schmidt modes are connected as follows:}
                    u_{n}^{\left(2\right)}\of{q} &= \E^{\frac{\I}{2}\phi} \psi_{n}^{\left(1\right)}\of{q}, 
                        \label{eq:conn_u2_psi1_phase} \\
                    \psi_{n}^{\left(2\right)}\of{q'} &= \E^{\frac{\I}{2}\phi} u_{n}^{\left(1\right)}\of{q'}.
                        \label{eq:conn_psi2_u1_phase}
                \end{align}
            \end{subequations}
          
            \par Using the connection relations~\cite{PRR5} 
            \begin{subequations}
                \begin{align}
                    \begin{split}
                        \beta^{\left(\mathrm{SU}\right)}\of{q,q'} &= \int\!\dd\bar{q}\,
                            \tilde{\eta}^{\left(2\right)}\of{q,\bar{q}} \beta^{\left(1\right)}\of{\bar{q},q'} \\
                        &\qquad+ \int\!\dd\bar{q}\,\beta^{\left(2\right)}\of{q,\bar{q}}
                            \left[\tilde{\eta}^{\left(1\right)}\of{\bar{q},q'}\right]^{*},
                            \label{eq:composite_tf_beta_nophase}
                    \end{split} \\
                    \begin{split}
                        \tilde{\eta}^{\left(\mathrm{SU}\right)}\of{q,q'} &= \int\!\dd\bar{q}\,
                            \tilde{\eta}^{\left(2\right)}\of{q,\bar{q}} \tilde{\eta}^{\left(1\right)}\of{\bar{q},q'} \\
                        &\qquad+ \int\!\dd\bar{q}\,\beta^{\left(2\right)}\of{q,\bar{q}}
                            \left[\beta^{\left(1\right)}\of{\bar{q},q'}\right]^{*},
                        \label{eq:composite_tf_eta_nophase}
                    \end{split}
                \end{align}
            \end{subequations}
            one can connect the transfer functions of the first and 
            the second crystal
            to obtain the transfer 
            functions~$\tilde{\eta}^{\left(\mathrm{SU}\right)}$
            and~$\beta^{\left(\mathrm{SU}\right)}$ of the whole SU(1,1) interferometer.
            
            Plugging the joint Schmidt decompositions for the
            transfer functions of the \revA{first and second crystal
            [see
            \Cref{eq:jointsdecompbeta,eq:jointsdecompeta}]}
            into \Cref{eq:composite_tf_beta_nophase,eq:composite_tf_eta_nophase},
            it becomes clear that
            the transfer functions 
            of the full SU(1,1) interferometer read
            \begin{subequations}
                \begin{align}
                    \begin{split}
                        \beta^{\left(\mathrm{SU}\right)}\of{q,q'} &= 
                            \sum_{n} \sqrt{\Lambda_{n}^{\left(\mathrm{SU}\right)}}
                            u_{n}^{\left(\mathrm{SU}\right)}\of{q}
                            \psi_{n}^{\left(\mathrm{SU}\right)}\of{q'},
                            \label{eq:sdecomp_su_vc1_betasu}
                    \end{split} \\
                    \begin{split}
                        \tilde{\eta}^{\left(\mathrm{SU}\right)}\of{q,q'} &=
                            \sum_{n} \sqrt{\tilde{\Lambda}_{n}^{\left(\mathrm{SU}\right)}}
                            u_{n}^{\left(\mathrm{SU}\right)}\of{q}
                            \left[\psi_{n}^{\left(\mathrm{SU}\right)}\of{q'}\right]^{*},
                    \end{split}
                \end{align}
                where
                \begin{align}
                    \Lambda_{n}^{\left(\mathrm{SU}\right)} &=
                        4 \cos^2\of{\frac{\phi}{2}}
                        \Lambda^{\left(1\right)}_n\tilde{\Lambda}^{\left(1\right)}_n, \\
                    \tilde{\Lambda}_{n}^{\left(\mathrm{SU}\right)} &= 
                        1 + \Lambda_{n}^{\left(\mathrm{SU}\right)}, \\
                    u_{n}^{\left(\mathrm{SU}\right)}\of{q} &= \exp\ofb{
                        \frac{\I}{2}\left(\mu+\zeta_n\right)}  \psi_{n}^{\left(1\right)}\of{q}, 
                        \label{eq:u_su_via_psi_1_comp} \\
                    \psi_{n}^{\left(\mathrm{SU}\right)}\of{q'} &= \exp\ofb{
                        \frac{\I}{2}\left(\mu-\zeta_n\right)}  \psi_{n}^{\left(1\right)}\of{q'},
                        \label{eq:psi_su_via_psi_1_comp}
                \end{align}
                with
                \begin{align}
                    \mu &= \arg\of{1+\E^{\I\phi}}, \\
                    \zeta_n &= \arg\ofb{1+\Lambda_n^{\left(1\right)}\left(1+\E^{\I\phi}\right)}.
                    \label{eq:def_zetan}
                \end{align}
            \end{subequations}
            Equations~(\ref{eq:sdecomp_su_vc1_betasu})%
            \crefrangeconjunction(\ref{eq:def_zetan})
            provide a joint Schmidt decomposition for the transfer
            functions of the entire interferometer
            in terms of the input Schmidt modes of the first
            crystal and supplement the discussion 
            of the Schmidt modes presented in \refcite{PRR5}.
            Furthermore, \Cref{eq:u_su_via_psi_1_comp,eq:psi_su_via_psi_1_comp} 
            suggest  that the input and output modes of the SU(1,1) interferometer
            are equal up to some constant phase. This is similar to the setup
            described in \refcite{10.1116/5.0203013} for the frequency domain,
            where the input and output modes of the system are also equal due to
            a certain symmetry in the system.

        \subsection{Unbalanced imperfectly compensated interferometer}%
                \label{sec:unbalanced_interf}

            For an unbalanced interferometer, the properties of the PDC
            section on the first and second pass are no longer identical, 
            meaning that the symmetry properties connecting the
            transfer functions of the first and second crystal described
            by \Cref{eq:conn_tf1tf2_beta,eq:conn_tf1tf2_eta} no longer
            hold and it is generally no longer possible to describe
            the connection between the transfer functions analytically.
            Thus, the transfer functions for both crystals have to be
            obtained by integrating the integro-differential
            equations numerically.

            \par For this kind of interferometer, we consider 
            phase matching functions of the form
            \begin{subequations}
                \begin{align}
                    h^{\left(1\right)}\of{q_s,q_i,L}
                        &= \E^{\I\Delta k L}, \label{eq:pm_func_c1_nonideal}\\
                    h^{\left(2\right)}\of{q_s,q_i,L}
                        &= \E^{-\I \Delta k\of{q_s,q_i}
                        \left[L-2L_1\right]}
                        \E^{-\I \Delta k^{\mathrm{air}}\of{q_s,q_i} \delta z}
                        \E^{\I\phi}.
                        \label{eq:h2_comp_imperfect}
                \end{align}
            \end{subequations}
            Compared to the balanced interferometer with full
            diffraction compensation described by \Cref{eq:h2_comp},
            the newly introduced phase term
            $\E^{-\I \Delta k^{\mathrm{air}}\of{q_s,q_i} \delta z}$
            in \Cref{eq:h2_comp_imperfect} describes an
            offset~$\delta z$ of the focusing element in the air gap of the
            SU(1,1) interferometer from its optimal
            position in the case of balanced gains.
            In experiments, this optimal position is normally 
            defined by the position that maximizes the 
            interferometric visibility 
            \begin{align}\label{eq:def_vis}
                v = \frac{\langle\hat{N}_{\mathrm{tot}}\rangle_{\mathrm{bf}}
                    - \langle\hat{N}_{\mathrm{tot}}\rangle_{\mathrm{df}}}
                {\langle\hat{N}_{\mathrm{tot}}\rangle_{\mathrm{bf}} 
                    + \langle\hat{N}_{\mathrm{tot}}\rangle_{\mathrm{df}}} \cdot\SI{100}{\percent},
            \end{align}
            when the phase is varied between the bright and dark fringe
            with integral
            intensities~$\langle\hat{N}_{\mathrm{tot}}\rangle_{\mathrm{bf}}$
            and~$\langle\hat{N}_{\mathrm{tot}}\rangle_{\mathrm{df}}$, respectively, 
            \revA{whereas}
            the phase $\phi$ is usually varied by
            changing the phase of the pump laser and is therefore independent 
            of $\delta z$, see also \Cref{fig:su11setup}.
            Clearly, for the perfectly compensated balanced interferometer,
            $v=\SI{100}{\percent}$, see
            \Cref{sec:app_vis} and \refcite{PRR5}.
            However, for
            an unbalanced (or imperfectly compensated)
            interferometer, the optimal visibility
            will generally be less than $\SI{100}{\percent}$.
             Generally, for a given setup, the optimal value for
            $\delta z$ can be found by varying $\delta z$ and
            maximizing the visibility, see \Cref{eq:form_vis_XY}, for which 
            an additional optimization over $\phi$ is not required.
            It should be noted that in general, as also described
            in \Cref{sec:app_vis},
            the bright and dark fringes do not occur at $\phi=0$
            and $\phi=\pi$, respectively and are instead shifted by some
            phase $\Upsilon$, see \Cref{eq:Upsilon}.

            \begin{figure*}%
                \centering%
                \newcommand{\densedots}{{.\kern-0.08em.\kern-0.08em.}}%
                \newcommand{\olcg}{\ocg}%
                \newcommand{\olch}{\och}%
                \newcommand{\lSU}{\left(\mathrm{SU}\right)}%
                \scalebox{1.2}{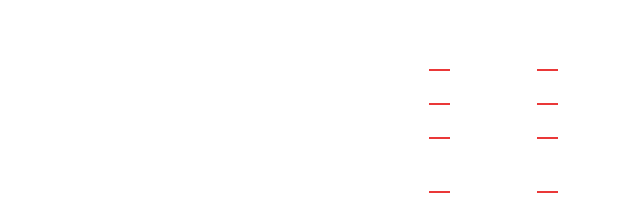}\\[1em]%
                \scalebox{1.2}{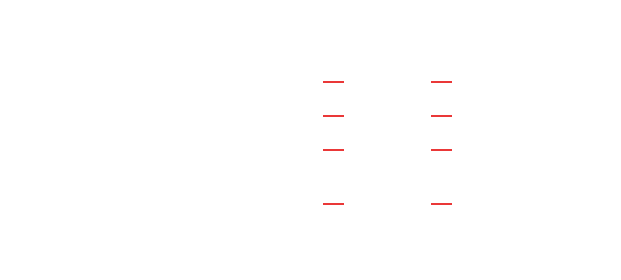}%
                \caption{\label{fig:modelayouts}%
                \revA{Upper row: Modal layout of the two crystals comprising
                the interferometers understood as separate PDC systems.
                The dashed lines drawn over the crystals illustrate the 
                Bogoliubov transformations connecting the input Schmidt modes
                $\psi_n^{\left(k\right)}$ to the output modes $u_n^{\left(k\right)}$
                with the same index $n$ (for $k=1,2$).
                The overlap coefficient $\ocg$ as defined in \Cref{eq:def_g_olap}
                describes the coupling between the
                output modes of the first crystal and the input modes of the 
                second crystal.
                Lower row: Modal layout of the SU(1,1) interferometer, with
                the modal layout of the two separate crystals shown in
                the background.
                This illustrates the main difference in the modes of the
                separate crystals and those of the interferometer:
                The modes associated with each crystal diagonalize their
                associated crystal, while the modes of the interferometer diagonalize
                the input-output relation of the interferometer.
                Notably, since $\ocg$ is in general not diagonal, an input-output
                relation connecting the $\psi_n^{\left(1\right)}$ to the
                $u_l^{\left(2\right)}$ will in general also not be diagonal and
                thus the $\psi_n^{\left(1\right)}$ and $u_l^{\left(2\right)}$
                may then not correspond to the input and output modes, respectively,
                of the entire interferometer.
                This explains the necessity to introduce the overlap coefficient
                $\och$ as defined in \Cref{eq:def_h_olap} in order to quantify
                the mismatch between the output modes of the second
                crystal $u_n^{\left(1\right)}$ and those of the interferometer
                $u_n^{\left(\mathrm{SU}\right)}$.
                Figure adapted from \refcite{diss}.}}
            \end{figure*}
            
            Before analyzing the behavior of the Schmidt
            modes in this kind of interferometer, we first
            \revA{illustrate the modal layouts
            of two crystals in comparison to the entire 
            interferometer in 
            \Cref{fig:modelayouts}. One can notice that to describe 
            the full SU(1,1) system, two sets of overlap 
            coefficients\footnote{In this work, we use Fraktur
            letters~$\ocg$ and~$\och$ (and~$\occ$)
            to designate the overlap coefficients which were
            dubbed $g$ and $h$ in \refcite{multimodesqueezing}
            in order to avoid confusion with the phase matching
            function~$h$ appearing in
            \Cref{eq:integ_diffeqs_eb_beta,eq:integ_diffeqs_eb_eta}.}
            $\ocg_{km}$ and $\och_{km}$ are required, which will be 
            introduced more formally below.}
            
            \revA{The first overlap coefficient corresponds to the 
            overlap between} the output modes of the
            first crystal and the input modes of the second
            crystal and is defined as~\cite{multimodesqueezing}:
            \begin{align}\label{eq:def_g_olap}
                \ocg_{km} &\eqdef \int\!\dd q
                    \left[u^{\left(1\right)}_k\of{q}\right]^{*}
                    \psi^{\left(2\right)}_m\of{q}.
            \end{align}
            In general, this overlap coefficient is not diagonal
            ($\ocg_{km}\not\propto\delta_{km}$), even for an
            interferometer consisting only of two consecutive
            crystals with no phase object in between, so
            that any output Schmidt mode of the first crystal may couple to
            all input Schmidt modes of the second crystal.
            
            \par In experimental setups, the modes that are usually
            measured are the \emph{output modes of the interferometer}
            and \emph{not} the output modes of the second crystal.
            These originate from the decomposition of the transfer functions
            of the entire interferometer and the transfer functions
            of the second crystal, respectively, and are, in general,
            not identical since they belong to different (sub-)systems.
            This leads to the definition of the overlap coefficients
            between the output modes of the second crystal and those
            of the entire interferometer~\cite{multimodesqueezing}:
            \begin{align}\label{eq:def_h_olap}
                \och_{nl} &\eqdef \int\!\dd q \left[u_n^{\left(2\right)}\of{q}\right]^{*}
                    u_l^{\left(\mathrm{SU}\right)}\of{q}.
            \end{align}
            \par Generally, both $\ocg_{km}$ and $\och_{nl}$ will
            depend on the interferometric phase since they depend
            on the Schmidt modes of the second crystal and
            the entire interferometer.

            \par
\begin{figure*}[p]%
                \centering%
                \includegraphics[width=\linewidth]{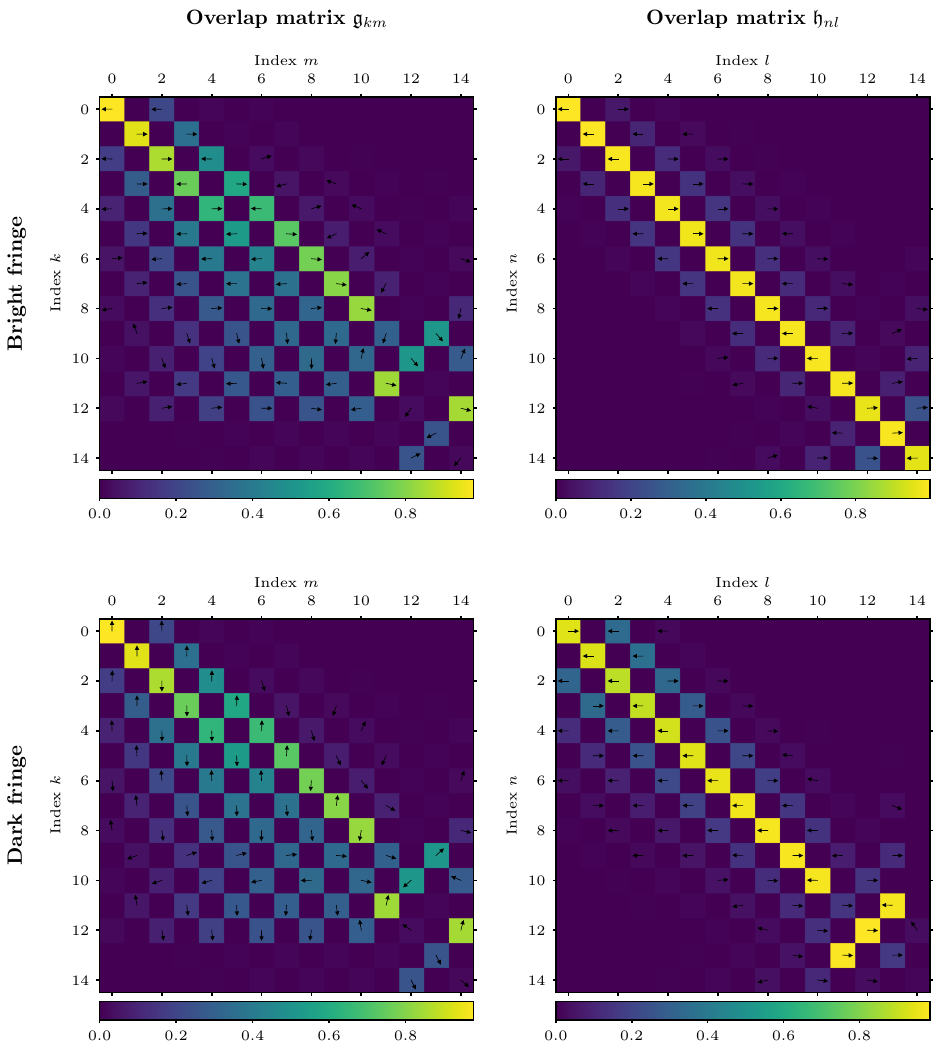}%
                \caption{\label{fig:omatsplots}%
                    First~$15\times 15$ entries of 
                    the overlap matrices~$\ocg_{km}$ [left column, defined in
                    \Cref{eq:def_g_olap}] and~$\och_{nl}$ [right column, defined in
                    \Cref{eq:def_h_olap}]
                    for the bright fringe (first row, $\phi=0.0337\pi$) and dark fringe
                    (second row, $\phi=1.03\pi$).
                    The color indicates the modulus of the complex-valued entry of the overlap
                    matrix and the handle pointing outwards from the center of each cell
                    indicates the location of the
                    value in the complex plane. That is, if the handle points to the right, the number
                    is purely real and positive and if it points up, the number is purely 
                    imaginary with positive sign. Handles are not shown for 
                    entries with modulus smaller than~$0.02$.
                    It should be noted that the phase of the overlap coefficients is only
                    defined up to adding or subtracting $\pi$ since the sign of the modes 
                    is not well-defined
                    from the joint Schmidt decomposition alone,
                    see \Cref{sec:uniq_sdecomp}.
                    See also \Cref{sec:app_vis}, which explains why the bright and dark
                    fringe do not occur exactly at $\phi=0$ and $\phi=\pi$,
                    respectively. Furthermore, \Cref{sec:gh_delta_z_phase} discusses the
                    alignment of the phases of the first few modes for varying $\delta z$.
                }%
            \end{figure*}

\begin{figure}%
                \centering%
                \includegraphics[width=\linewidth]{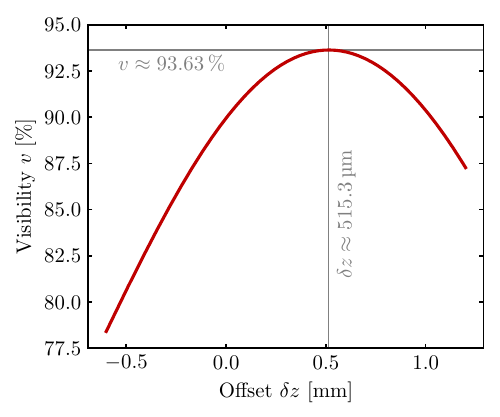}%
                \caption{\label{fig:visoverdeltaz}%
                    Visibility~$v$ of an unbalanced SU(1,1)
                    interferometer with parametric
                    gains~$G^{\left(1\right)}=1$ and~$G^{\left(2\right)}=4$
                    for the first and second crystal, respectively,
                    and the remaining parameters
                    as in \Cref{sec:single_crystal}, as a function of the 
                    offset~$\delta z$. The visibility~$v$ reaches it maximum
                    value of~$\SI{93.63}{\percent}$ at
                    an offset of~$\delta z\approx\SI{515.3}{\micro\metre}$,
                    relative to the optimal position of a perfectly compensated
                    interferometer with equal gains in both crystals ($\delta z = 0$).
                }%
            \end{figure}

 Next, in \Cref{fig:omatsplots}, 
            we provide examples for the overlap
            matrices~$\ocg$ and~$\och$ defined above, where
            the first $15\times 15$ entries of the overlap matrices
            are shown.
            For this, based on \refcite{multimodesqueezing},
            we consider an unbalanced SU(1,1) interferometer
            with gains~$G^{\left(1\right)}_{\mathrm{exp}}=1$ 
            and~$G^{\left(2\right)}_{\mathrm{exp}}=4$
            for the first and second crystal, respectively.
            The remaining parameters are the same as in the
            case of the single crystal discussed above, see
            \Cref{sec:single_crystal}.
            \revA{For this setup, the dependence of interferometric visibility $v$ as defined in 
            \Cref{eq:def_vis} on $\delta z$ can be seen in
            \Cref{fig:visoverdeltaz} and takes it maximum value if the focusing element
            inside the interferometer
            has an offset of~$\delta z=\SI{515.3}{\micro\metre}$ from
            its optimal position in the case of balanced interferometer.
            Note that generally the optimal value of $\delta z$ will
            depend on the system parameters.}

            \par From the plots
            of~$\ocg$ in \Cref{fig:omatsplots} it is clear that the output modes of the first crystal
            and the input modes of the second crystal are substantially different,
            leading to the checkerboard-pattern visible in the lower left half.
            It should be noted that when comparing~$\ocg$ for the bright and dark
            fringes, only the phases of the matrix entries are affected. 
            This can
            be seen as follows: As was shown in \refcite{PRR5}, for a PDC section
            described by the phase matching functions of the form
            as in \Cref{eq:h2_comp,eq:h2_comp_imperfect}, it is possible to analytically
            account for the flat phase contribution $\E^{\I\phi}$ by using the fact
            that the transfer function $\beta$ can be written in the form~\cite{PRR5}
            \begin{align}
                \beta\of{q_s,q_i} = \E^{\I\phi} \nophase{\beta}\of{q_s,q_i}
            \end{align}
            where $\nophase{\beta}\of{q_s,q_i}$ is the transfer function
            resulting from the integro-differential equations by integrating them without
            the phase contribution $\E^{\I\phi}$ in the phase matching function, or,
            equivalently, for $\phi=0$.
            Generally, in the following, the notation
            \begin{align}\label{eq:nophase_notation}
                \nophase{x}
            \end{align}
            will be used to denote a quantity $x$ obtained by solving
            the integro-differential equations with the 
            replacement $\E^{\I\phi}\rightarrow 1$.
            The transfer function $\tilde{\eta}$ does not change when removing
            the phase contribution~\cite{PRR5}, but the notation introduced
            above may still be applied here, since $\beta$ and
            $\tilde{\eta}$ are obtained as a pair when solving the
            integro-differential equations.
            The joint decomposition of the transfer functions
            for the second crystal
            can then be written in a form analogous 
            to \Cref{eq:sdecomp_beta,eq:sdecomp_eta}:
            \begin{subequations}
                \begin{align}
                    \beta^{\left(2\right)}\of{q,q'} &= \sum_{n} \sqrt{\Lambda_{n}^{\left(2\right)}}
                        \left[\E^{\frac{\I}{2}\phi}
                        \nophase{\psi}_{n}^{\left(2\right)}\of{q}\right]
                        \left[\E^{\frac{\I}{2}\phi} 
                        \nophase{u}_{n}^{\left(2\right)}\of{q'}\right],
                        \label{eq:sdecomp_beta_imperfect} \\
                    \tilde{\eta}^{\left(2\right)}\of{q,q'} &= \sum_{n} \sqrt{\tilde{\Lambda}_{n}^{\left(2\right)}}
                        \left[\E^{\frac{\I}{2}\phi} 
                            \nophase{\psi}_{n}^{\left(2\right)}\of{q}\right]
                        \left[\E^{\frac{\I}{2}\phi}
                            \nophase{u}_{n}^{\left(2\right)}\of{q'}\right]^{*},
                        \label{eq:sdecomp_eta_imperfect}
                \end{align}
            \end{subequations}
            where the full dependence on the phase is given by
            the terms~$\E^{\frac{\I}{2}\phi}$ and the functions $\nophase{u}_{n}$
            are phase independent, as described for $\nophase{\beta}\of{q_s,q_i}$
            above.
            Thus, to explicitly account for the phase, 
            the defining expression for $\ocg$
            as given in
            \Cref{eq:def_g_olap} may be rewritten as
            \begin{align}\label{eq:g_with_phase_imperfect}
                \ocg_{km} &= \E^{\I\frac{\phi}{2}}\int\!\dd q
                    \left[u^{\left(1\right)}_k\of{q}\right]^{*}
                    \nophase{\psi}^{\left(2\right)}_m\of{q},
            \end{align}
            where~$\nophase{\psi}^{\left(2\right)}_m\of{q}$ now refers to the
            phase-independent input mode of the second crystal.
            Importantly however, note that in this section,
            \Cref{eq:conn_tf1tf2_beta,eq:conn_tf1tf2_eta}
            no longer hold due to the fact that both crystals have different
            gain values\footnote{%
                Even if both crystals have the same parametric gain,
                the~$q_s$- and~$q_i$-dependent phase 
                term~$\E^{-\I \Delta k^{\mathrm{air}}\of{q_s,q_i} \delta z}$
                in \Cref{eq:h2_comp_imperfect} still prevents this. 
            }
            and thus \Cref{eq:sdecomp_beta_imperfect,eq:sdecomp_eta_imperfect}
            cannot be trivially 
            expressed in terms of the singular values and mode
            functions of the first crystal, as was the case in
            \Cref{eq:sdecomp_beta,eq:sdecomp_eta}.

            \par In \Cref{eq:g_with_phase_imperfect}, the integral does not 
            depend on the interferometer phase~$\phi$.
            Hence, the phase dependence is fully described by the
            term~$\E^{\I\frac{\phi}{2}}$, which explains why the modulus
            of~$\ocg$ does not change as the phase varies from the dark fringe to
            the bright fringe. Furthermore, this also implies
            that the phase of~$\ocg$ differs by exactly~$\pi$ when comparing
            the bright and dark fringe, apart from additional $\pi$-phase jumps due
            to the fact that the sign of the Schmidt modes obtained
            from the numerical decomposition is arbitrary, see \Cref{sec:uniq_sdecomp}.

            \par Note that for the balanced case with perfect
            compensation, the overlap coefficients
            are trivially diagonal:
            \begin{subequations}
                \begin{align}
                    \ocg_{km} &= \E^{\I\frac{\phi}{2}} \delta_{km},
                \end{align}
                compare \Cref{eq:def_g_olap,eq:conn_psi2_u1_phase}, and
                \begin{align}
                    \och_{ln} &= \E^{\frac{\I}{2}\left(\mu+\zeta_n-\frac{\phi}{2}\right)} \delta_{ln},
                \end{align}
            \end{subequations}
            see \Cref{eq:def_h_olap,eq:u_su_via_psi_1_comp,eq:conn_u2_psi1_phase}.
            For numerical calculations, as mentioned above, it must be taken
            into account that random phase jumps may lead to sign changes
            of the overlap coefficients.

            \par \revA{The observations made above raise questions regarding
            the physical interpretations of the Schmidt modes. Care
            must be taken to not confuse the Schmidt modes of the two
            isolated crystals of the interferometers (that diagonalize the solution describing each separate crystal) and the Schmidt modes
            of the composite system (that diagonalize the full two-crystal system), as depicted  in
            \Cref{fig:modelayouts}.}
            
            \par \revA{In this context, it is also important
            to address a common misconception:
            While the term
            \enquote{input} Schmidt mode implies that it is
            connected to the input plane-wave operators, which correspond to the
            vacuum operators for the \emph{first} PDC section, their shapes
            $\psi_n$, as well as the shapes of the output modes $u_n$,
            depend on the physical parameters of the PDC process
            (the length of the PDC section,
            its refractive indices, the pump profile, the parametric gain).
            In more direct terms, it is not possible to obtain these modes 
            without having computed the transfer functions $\tilde{\eta}$
            and $\beta$ for the entire interaction, since only 
            after computing them it is possible to
            find the set of modes which diagonalizes the plane-wave input/output
            relations as written in \Cref{eq:as_dagger_sol,eq:ai_dagger_sol}.
            More precisely: The input modes are \emph{not} known \enquote{before}
            the interaction and are \emph{not} a property of the fields as they appear
            before the interaction at the input faces of the crystals, but rather
            a property of the PDC interaction.~\cite{diss}}

            \par \revA{Note that the description of
            the gain-unbalanced interferometer as given in this section can
            also  be generalized to SU(1,1) interferometers with different
            crystal lengths,
            refractive indices and other differing properties of the
            two crystals.}

    \section{MULTIMODE SQUEEZING MEASUREMENT}\label{sec:mmsqm}
       
        \revA{The modal structure of the SU(1,1) 
        interferometer described above can be used for measuring
        the level of squeezing in the $n$th output mode
        $u_n^{\left(1\right)}\of{q}$ of the first crystal based 
        on measurements of properties of the 
        second crystal and the full SU(1,1) interferometer 
        (phase sensitive parametric amplification).
        The simplified version of this technique was 
        experimentally realized in \refcite{multimodesqueezing}.
        In this section, we present a rigorous mathematical 
        derivation of this technique and discuss simplifications 
        that should be provided to make this technique experimentally feasible.}
        
        \par \revA{First, it should be highlighted that since the ultimate goal
        is to measure the squeezing in the modes associated with the mode
        functions $u_n^{\left(1\right)}$ of the first crystal, it is necessary
        to split the entire interferometer into this crystal and the following
        crystal. Otherwise, it is not possible to access these modes
        $u_n^{\left(1\right)}$, as can also be seen by careful 
        inspection of \Cref{fig:modelayouts}.}
        For the unbalanced interferometer, due to the non-diagonal overlap 
        matrices $\ocg$ and $\och$, the information regarding this mode
        $u_n^{\left(1\right)}\of{q}$
        is distributed over all output modes 
        $u_k^{\left(\mathrm{SU}\right)}\of{q}$ of the entire 
        interferometer\revA{, which in turn contains a mixture of 
        the information of all modes $u_n^{\left(1\right)}\of{q}$.} However,
        provided that the overlap coefficients and the eigenvalues
        of the second crystal and the interferometer are known,
        it is possible to reconstruct all information regarding
        the $u_n^{\left(1\right)}\of{q}$.
        Similarly, it is possible to connect the output
        Schmidt operators $\hat{A}_l^{\left(1,\mathrm{out}\right)}$
        of the first crystal to the output operators 
        $\hat{A}_m^{\left(\mathrm{SU},\mathrm{out}\right)}$
        of the entire interferometer, as will be seen in the following.

        \subsection{Exact reconstruction procedure}%
                \label{sec:exact_reconstruction}
                
            \par When connecting the first crystal of the interferometer
            to the second crystal, the underlying assumption 
            is that the output plane-wave operators of the first crystal
            coincide with the input plane-wave operators of the
            second crystal:
            \begin{align}\label{eq:io_pw12}
                \hat{a}^{\left(1,\mathrm{out}\right)}\of{q} &= \hat{a}^{\left(2,\mathrm{in}\right)}\of{q}.
            \end{align}
            This assumption leads also to
            the connection relations connecting
            the transfer functions of the two crystals
            to those of the entire interferometer
            as written in
            \Cref{eq:composite_tf_beta_nophase,%
            eq:composite_tf_eta_nophase}~\cite{PRR5}.
            The relation for the corresponding connection between the 
            Schmidt mode operators is not as simple as \Cref{eq:io_pw12}.
            Instead, using \Cref{eq:def_g_olap,eq:io_pw12}, one finds
            \begin{align}\label{eq:A1out_gA2in}
                \hat{A}_l^{\left(1,\mathrm{out}\right)} 
                    &= \sum_{m} \ocg_{lm} \hat{A}_m^{\left(2,\mathrm{in}\right)}.
            \end{align}
            The input Schmidt operators of the second crystal are
            transformed through the second crystal using the
            Bogoliubov transforms. The inverse transform for
            \Cref{eq:bogoliubov_plain} reads\footnote{This follows
                for example similarly to \Cref{eq:bogoliubov_plain}
                from Eqs.~(D2a) and (D2b) of \refcite{PRR5}
                and can be verified by combining
                \Cref{eq:bogoliubov_plain,eq:bogoliubov_plain_inverse}
                and using the connections between the
                eigenvalues as written in \Cref{eq:connection_L_Ltilde}.%
            }:
            \begin{align}\label{eq:bogoliubov_plain_inverse}
                \hat{A}_m^{\left(\mathrm{in}\right)} &= 
                \sqrt{\tilde{\Lambda}_m}\,\hat{A}_m^{\left(\mathrm{out}\right)}
                - \sqrt{\Lambda_m}\left[\hat{A}_m^{\left(\mathrm{out}\right)}\right]^{\dagger}.
            \end{align}
            Plugging this into \Cref{eq:A1out_gA2in} yields an 
            expression connecting the output operators of the
            first crystal with the
            output operators of the second crystal.
            \par At the output of the second crystal,
            the output plane-wave operators of the interferometer
            coincide with the output plane-wave operators of the
            second crystal: $\hat{a}^{\left(2,\mathrm{out}\right)}\of{q} 
            = \hat{a}^{\left(\mathrm{SU},\mathrm{out}\right)}\of{q}$.
            However, this does, again, not hold analogously for
            the Schmidt mode operators. Instead, analogously to
            \Cref{eq:io_pw12} one finds:
            \begin{align}\label{eq:io_pw2su}
                \hat{A}_m^{\left(2,\mathrm{out}\right)} 
                    &= \sum_{k} \och_{mk} \hat{A}_k^{\left(\mathrm{SU},\mathrm{out}\right)}.
            \end{align}
            \par Finally, combining
            \Cref{eq:A1out_gA2in,eq:bogoliubov_plain_inverse,eq:io_pw2su}
            as detailed above, we obtain
            the following expression connecting the output Schmidt operators
            of the first crystal with those of 
            the entire interferometer:
            \begin{align}\label{eq:A1out_via_suout}
                \begin{split}
                    \hat{A}_l^{\left(1,\mathrm{out}\right)} 
                        &= \sum_{mk} \ocg_{lm} \bigg[
                            \sqrt{\tilde{\Lambda}^{\left(2\right)}_m}
                            \och_{mk}\,\hat{A}_k^{\left(\mathrm{SU},\mathrm{out}\right)} \\
                        &\qquad- \sqrt{\Lambda_m^{\left(2\right)}}\och_{mk}^{*}
                            \left(\hat{A}_k^{\left(\mathrm{SU},\mathrm{out}\right)}\right)^{\dagger}
                        \bigg].
                \end{split}
            \end{align}
            \par The \emph{generic 
            quadrature} operator for the $l$th Schmidt mode can be 
            defined analogously to the plane-wave quadrature
            operators~\cite{Gerry_Knight_2004} as
            \begin{align}\label{eq:gen_quad_def}
                \hat{P}_{\vartheta,l} &\eqdef \hat{A}_l\E^{-\I\frac{\vartheta}{2}} + 
                    \hat{A}_l^{\dagger}\E^{\I\frac{\vartheta}{2}},
            \end{align}
            where $\vartheta$ is the phase of the generic quadrature.
            For $\vartheta=\pi$ and $\vartheta=0$, the following special
            quadrature operators can be defined:
            \begin{subequations}
                \begin{align}
                    \hat{X}_l \eqdef \hat{P}_{\pi,l} &= -\I \left(\hat{A}_l - \hat{A}_l^{\dagger}\right),
                        \label{eq:X_op_A}\\
                    \hat{Y}_l \eqdef \hat{P}_{0,l}  &= \hat{A}_l + \hat{A}_l^{\dagger}, \label{eq:Y_op_A}
                \end{align}
            \end{subequations}
            which are analogous to the quadrature operators 
            defined during the quantization procedure,
            see for example \refcite{Gerry_Knight_2004,Scully_Zubairy_1997}.
            Generally, the generic quadrature operators fulfill the 
            following commutation relation:
            \begin{align}
                \comm{\hat{P}_{\vartheta+\pi,l}}{\hat{P}_{\vartheta,l}^{\dagger}} &= \frac{2}{\I}.
            \end{align}
            For squeezed vacuum, $ \langle \hat{P}_{\vartheta,l}\rangle=0 $, and the 
            variance\footnote{%
                In this work,
                we use the notation $\Dvar$ for the variance of an operator: 
                $\Dvar\hat{X} = \langle\hat{X}^2\rangle-\langle\hat{X}\rangle^2$.
            } of the generic quadrature operators is given by
            \begin{align}\label{eq:genPquadvar}
                \Dvar \hat{P}_{\vartheta,l} = 1+ 2 \langle \hat{A}^{\dagger}_l \hat{A}_l \rangle 
                    + 2\Re\ofb{\langle \hat{A}_l   \hat{A}_l \rangle\,\E^{-\I\vartheta}}, 
            \end{align}
            which reduces to
            \begin{subequations}
                \begin{align}
                    \Dvar \hat{X}_l = 1+ 2 \langle \hat{A}^{\dagger}_l \hat{A}_l \rangle - 2\Re\ofb{\langle \hat{A}_l \hat{A}_l \rangle}, 
                        \label{eq:X_Quadrature_var}\\
                    \Dvar \hat{Y}_l = 1+ 2 \langle \hat{A}^{\dagger}_l \hat{A}_l \rangle + 2\Re\ofb{\langle \hat{A}_l \hat{A}_l \rangle},
                        \label{eq:Y_Quadrature_var}
                \end{align}
            \end{subequations}
            for the special case written in \Cref{eq:X_op_A,eq:Y_op_A}.

            \par It should be noted that \Cref{eq:genPquadvar}
            implies that the concrete value for $\vartheta$
            to reach the maximally squeezed and anti-squeezed quadrature may be shifted
            away from $\vartheta=\pi$ and $\vartheta=0$, respectively, if
            $\langle \hat{A}_l \hat{A}_l \rangle$ is complex-valued.
            Generally, the maximal levels of squeezing 
            and anti-squeezing are defined
            as the minimum and maximum of $\Dvar \hat{P}_{\vartheta,l}$, respectively.
            These extremal values for the quadrature variance
            may be written as
            \begin{subequations}
                \begin{align}
                    \min_{\vartheta}  \Dvar \hat{P}_{\vartheta,l} 
                        = \Dvar \hat{P}_{\vartheta_{\mathrm{min},l},l} 
                        &= 1+ 2 \langle \hat{A}^{\dagger}_l \hat{A}_l \rangle 
                            - 2 \abs{\langle \hat{A}_l   \hat{A}_l \rangle},
                            \label{eq:min_P_theta_res}\\
                    \max_{\vartheta}  \Dvar \hat{P}_{\vartheta,l} 
                        = \Dvar \hat{P}_{\vartheta_{\mathrm{max},l},l} 
                        &= 1+ 2 \langle \hat{A}^{\dagger}_l \hat{A}_l \rangle 
                            + 2 \abs{\langle \hat{A}_l   \hat{A}_l \rangle},
                            \label{eq:max_P_theta_res}
                \end{align}
            \end{subequations}
            where we have introduced $\vartheta_{\mathrm{min},l}$
            and $\vartheta_{\mathrm{max},l}$ as the values for $\vartheta$
            that minimize and maximize the quadrature variance, respectively.
            With this, the levels of squeezing ($\Sq$) and
            anti-squeezing ($\Asq$) in mode $l$ are given by:
            \begin{subequations}
                \begin{align}
                    \Sq_l &= 10 \log_{10}\ofb{
                    \frac{\min_{\vartheta} \Dvar\hat{P}_{\vartheta,l}}{\Dvar \hat{P}^{\left(\mathrm{vac}\right)}}} 
                        \unit{\deci\bel},
                        \label{eq:sq_def} \\
                    \Asq_l &= 10 \log_{10}\ofb{\frac{\max_{\vartheta}\Dvar\hat{P}_{\vartheta,l}}{\Dvar \hat{P}^{\left(\mathrm{vac}\right)}}}
                        \unit{\deci\bel}, \label{eq:asq_def}
                \end{align}
            \end{subequations}
            where $\hat{P}^{\left(\mathrm{vac}\right)}$
            is the corresponding vacuum quadrature operator,
            which has a quadrature variance independent of
            $\vartheta$, which is why the corresponding index
            was dropped.
            To be more precise, from \Cref{eq:genPquadvar}
            it is immediately clear that
            \begin{align}\label{eq:vac_quad_1}
                \Dvar\hat{P}^{\left(\mathrm{vac}\right)}=1.
            \end{align}
            According to the definitions written in \Cref{eq:sq_def,eq:asq_def} 
            the levels of squeezing and anti-squeezing
            are measures of the relative change in quadrature
            variance compared to the vacuum quadrature variance.

            \par Equations (\ref{eq:genPquadvar}), (\ref{eq:sq_def}) 
            and (\ref{eq:asq_def}) show that the levels of squeezing
            and anti-squeezing can be obtained from the expectation
            values $\langle \hat{A}^{\dagger}_l \hat{A}_l \rangle$
            and $\langle \hat{A}_l \hat{A}_l \rangle$.
            By forming these expectation values for the output Schmidt
            mode operators of the first crystal as it is written in
            \Cref{eq:A1out_via_suout}, it is possible to express the
            quadrature variance and thereby also the levels of squeezing
            and anti-squeezing via the overlap matrices $\ocg$ and $\och$ and 
            the eigenvalues of the second crystal and the entire interferometer.
            The concrete expressions for the expectation values read:            
            \begin{widetext}
                \begingroup
                \allowdisplaybreaks
                \begin{subequations}
                    \begin{align}
                        \begin{split}\label{eq:exact_reco_AA}
                            \langle\hat{A}_l^{\left(1,\mathrm{out}\right)}\hat{A}_l^{\left(1,\mathrm{out}\right)}\rangle 
                            &= \sum_{n,n',k} \ocg_{ln} \ocg_{ln'} 
                                \Bigg[\och_{nk} \och_{n'k} \sqrt{\left(\Lambda^{\left(2\right)}_n+1\right) 
                                    \left(\Lambda^{\left(2\right)}_{n'}+1\right)}
                                    \sqrt{\left(\Lambda^{\left(\mathrm{SU}\right)}_{k}+1\right)
                                        \Lambda^{\left(\mathrm{SU}\right)}_{k}} \\
                            &\qquad+\och^*_{nk} \och^*_{n'k} \sqrt{\Lambda^{\left(2\right)}_n \Lambda^{\left(2\right)}_{n'} }
                                \sqrt{\Lambda^{\left(\mathrm{SU}\right)}_{k} \left(\Lambda^{\left(\mathrm{SU}\right)}_{k}+1\right) } \\
                            &\qquad- \och_{nk} \och^*_{n'k} \sqrt{\left(\Lambda^{\left(2\right)}_n+1\right) \Lambda^{\left(2\right)}_{n'} }
                                \left(\Lambda^{\left(\mathrm{SU}\right)}_{k}+1\right) \\
                            &\qquad- \och^*_{nk} \och_{n'k} \sqrt{\Lambda^{\left(2\right)}_n \left(\Lambda^{\left(2\right)}_{n'}+1\right) } 
                                \Lambda^{\left(\mathrm{SU}\right)}_k 
                           \Bigg],
                        \end{split}
                    \shortintertext{and}
                        \begin{split}\label{eq:exact_reco_AdA}
                            \langle\left(\hat{A}_l^{\left(1,\mathrm{out}\right)}\right)^{\dagger} 
                                \hat{A}_l^{\left(1,\mathrm{out}\right)} \rangle 
                            &= \sum_{n,n',k} \ocg^*_{ln} \ocg_{ln'}
                                \Bigg[\och^*_{nk} \och_{n'k} \sqrt{\left(\Lambda^{\left(2\right)}_n+1\right) 
                                    \left(\Lambda^{\left(2\right)}_{n'}+1\right)} 
                                    \Lambda^{\left(\mathrm{SU}\right)}_{k} \\
                            &\qquad+ \och_{nk} \och^*_{n'k} \sqrt{\Lambda^{\left(2\right)}_n 
                                \Lambda^{\left(2\right)}_{n'} }
                                \left(\Lambda^{\left(\mathrm{SU}\right)}_{k}+1\right) \\
                            &\qquad- \och^*_{nk} \och^*_{n'k} 
                                \sqrt{\left(\Lambda^{\left(2\right)}_n+1\right) 
                                \Lambda^{\left(2\right)}_{n'} } 
                                \sqrt{\Lambda^{\left(\mathrm{SU}\right)}_k 
                                    \left(\Lambda^{\left(\mathrm{SU}\right)}_{k}+1\right) } \\
                            &\qquad- \och_{nk} \och_{n'k} \sqrt{\Lambda^{\left(2\right)}_n
                                \left(\Lambda^{\left(2\right)}_{n'}+1\right)}
                                \sqrt{\Lambda^{\left(\mathrm{SU}\right)}_k 
                                    \left(\Lambda^{\left(\mathrm{SU}\right)}_{k}+1\right) } 
                            \Bigg].
                        \end{split}
                    \end{align}
                \end{subequations}
                \endgroup
            \end{widetext}
            In order to evaluate these expectation values, we have used
            the Bogoliubov transforms of the Schmidt operators from 
            \Cref{eq:bogoliubov_plain} written in terms of
            the entire interferometer to connect its output operators
            appearing in \Cref{eq:A1out_via_suout}, with its input operators.
            The contributing expectation values read
            \begin{subequations}
                \begin{align}
                    \begin{split}
                        \langle \hat{A}_l^{\left(\mathrm{out},\mathrm{SU}\right)}
                            \hat{A}_l^{\left(\mathrm{out},\mathrm{SU}\right)}\rangle
                            &= \langle \left(\hat{A}_l^{\left(\mathrm{out},\mathrm{SU}\right)}\right)^{\dagger}
                            \left(\hat{A}_l^{\left(\mathrm{out},\mathrm{SU}\right)}\right)^{\dagger}\rangle\\
                            &= \sqrt{\left(1+\Lambda_l^{\left(\mathrm{SU}\right)}\right)
                                \Lambda_l^{\left(\mathrm{SU}\right)}}, \label{eq:expval_AA}
                    \end{split}\\
                    \begin{split}
                    \langle \left(\hat{A}_l^{\left(\mathrm{out},\mathrm{SU}\right)}\right)^{\dagger}
                        \hat{A}_l^{\left(\mathrm{out},\mathrm{SU}\right)}\rangle
                        &= \langle \hat{A}_l^{\left(\mathrm{out},\mathrm{SU}\right)}
                        \left(\hat{A}_l^{\left(\mathrm{out},\mathrm{SU}\right)}\right)^{\dagger}\rangle - 1\\
                        &= \Lambda_l^{\left(\mathrm{SU}\right)}.  \label{eq:expval_AdA}
                    \end{split}
                \end{align}
            \end{subequations}
            \par Thus, given Eqs. (\ref{eq:genPquadvar}) and
            (\ref{eq:min_P_theta_res})\crefrangeconjunction(\ref{eq:expval_AdA})
            and complete knowledge regarding the quantities
            $\ocg_{ln}$, $\och_{nk}$, $\Lambda^{\left(2\right)}_n$ and
            $\Lambda^{\left(\mathrm{SU}\right)}_k$,
            it is possible to compute the levels of squeezing 
            and anti-squeezing in each output mode of the first crystal.

        \subsection{High-gain amplifier approximation}%
                \label{sec:high_gain_approx_reconstruction}
            The exact expressions for the quadrature variance resulting from
            \Cref{eq:exact_reco_AA,eq:exact_reco_AdA} are rather complex.
            Furthermore, it requires knowledge of the eigenvalues $\Lambda_n$,
            referring to absolute photon numbers in the modes, which are quite 
            tricky to obtain in experiments. 
            Instead, the relative intensities between modes are free from
            setup-dependent prefactors and are therefore much easier to extract.
            To address this, in the following, we will derive
            an approximated method for
            obtaining the squeezing in each mode which
            ultimately only relies on the
            \emph{relative} values of the eigenvalues.

            \begin{figure*}[!p]%
                \centering%
                \subfloat{\includegraphics[width=0.5\linewidth]{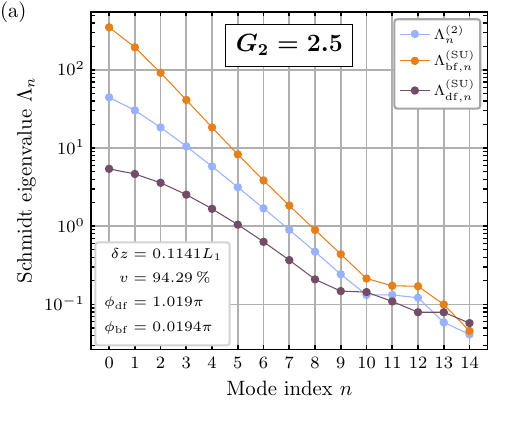}}%
                \subfloat{\includegraphics[width=0.5\linewidth]{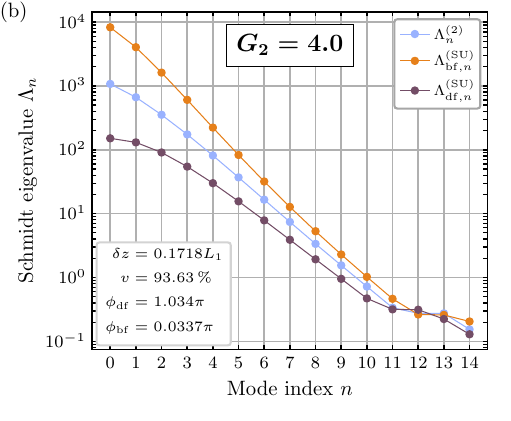}}\\
                \subfloat{\includegraphics[width=0.5\linewidth]{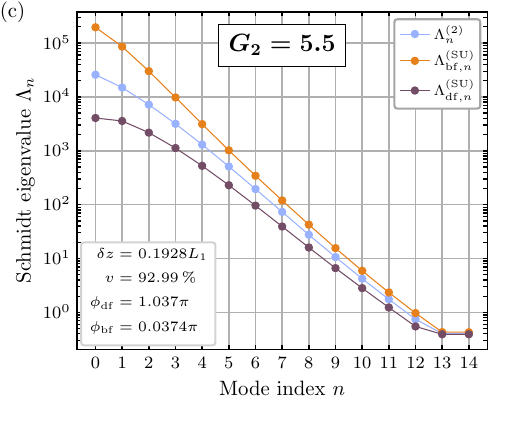}}%
                \subfloat{\includegraphics[width=0.5\linewidth]{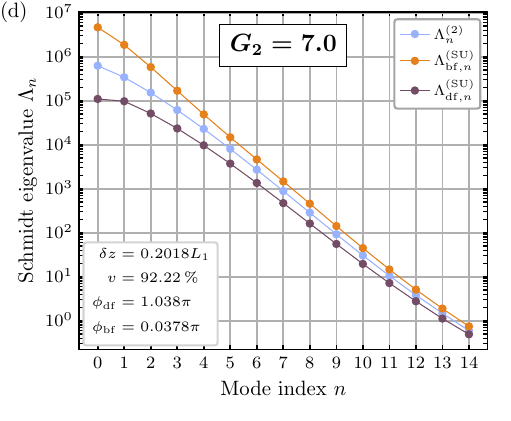}}%
                \caption{\label{fig:eigvals}%
                    Eigenvalues $\Lambda_n$ for the first 15 Schmidt modes
                    of the amplifier (second crystal) 
                    and the entire SU(1,1) interferometer at the dark and bright fringe
                    for different gain values as written in the plots. The 
                    corresponding values for
                    $\delta z$ and the visibility $v$, as well as the
                    phases $\phi_{\mathrm{df}}$ 
                    and $\phi_{\mathrm{bf}}$ for the dark and bright fringes, respectively,
                    are also provided in the plots. The lines connecting
                    the data points serve as a visual guide. Clearly,
                    $\Lambda_{\mathrm{df},n}^{\left(\mathrm{SU}\right)}
                    <\Lambda_{n}^{\left(2\right)}
                    <\Lambda_{\mathrm{bf},n}^{\left(\mathrm{SU}\right)}$
                    for the first few modes $n<k$, where $k$ increases as
                    the parametric gain is increased. The condition
                    $\Lambda_{\mathrm{df},n}^{\left(\mathrm{SU}\right)} \gg 1$
                    holds true approximately
                    up to $n=5$ in (a), for $n=9$ in (b), for $n=11$ 
                    in (c) and for $n=13$ in (d), meaning that as the gain
                    increases, the high-gain amplifier approximation
                    holds up to higher mode indices $n$.
                }%
            \end{figure*}

            \par If the gain of the second crystal (amplifier)
            is large enough for the inequalities
            $\Lambda^{\left(2\right)}_n \gg 1$, 
            $\Lambda^{\left(\mathrm{SU}\right)}_n \gg 1$ to be satisfied
            (see \Cref{fig:eigvals} for plots of the eigenvalues),
            it is possible to apply the following approximations
            for the eigenvalues of the $\tilde{\eta}$ transfer function
            of the second crystal and the entire 
            interferometer\footnote{These
            can be obtained as follows: An expression of the form
            $\sqrt{x+1}$ can be factorized as $\sqrt{x}\sqrt{1+1/x}$.
            For large $x$, the second square root can be expanded 
            into its Taylor series (binomial series) for $x>1$,
            since $1/x<1$. Thus, $\sqrt{1+1/x}=1+1/\!\left(2x\right)+\OO\of{1/x^2}$.
            Truncating the
            series after the second term leads to the binomial approximation
            $\sqrt{1+1/x}\approx 1+1/\!\left(2x\right)$. Plugging this back
            into the original expression directly leads to the approximations
            as written in \Cref{eq:approx_L2,eq:approx_LSU}.}:
            \begin{subequations}
                \begin{align}
                     \sqrt{\tilde{\Lambda}^{\left(2\right)}_{n}} = \sqrt{\Lambda^{\left(2\right)}_{n}+1} &\approx 
                        \sqrt{\Lambda^{\left(2\right)}_{n}} \left(1 + \frac{1}{2 \Lambda^{\left(2\right)}_{n}}\right),~ \label{eq:approx_L2} \\
                     \sqrt{\tilde{\Lambda}^{\left(\mathrm{SU}\right)}_{k}} = \sqrt{\Lambda^{\left(\mathrm{SU}\right)}_{k}+1} &\approx \sqrt{\Lambda^{\left(\mathrm{SU}\right)}_{k}} 
                        \left(1 + \frac{1}{2 \Lambda^{\left(\mathrm{SU}\right)}_{k}}\right).
                  \label{eq:approx_LSU}
                \end{align}
            \end{subequations}
            
            It should be noted that $\Lambda^{\left(2\right)}_n \gg 1$ is 
            satisfied by the assumption that the amplifier operates at high gain. 
            However, it is not immediately obvious that
            $\Lambda^{\left(\mathrm{SU}\right)}_n \gg 1$ holds in general,
            since the output intensity of the interferometer
            should usually be small at the dark fringe \revA{(relative to the 
            average intensity as the phase varies)}. 
            Nevertheless, if the gain of the amplifier is much
            larger than the gain of the squeezer (the first crystal), 
            meaning  $G^{\left(2\right)}\gg G^{\left(1\right)}$,
            the eigenvalues of the interferometer are reasonably large 
            even at the dark fringe, see \Cref{fig:eigvals}. 
            \par Applying the approximations made in 
            \Cref{eq:approx_L2} and \Cref{eq:approx_LSU} to 
            \Cref{eq:A1out_via_suout}, the output Schmidt operator of
            the first crystal can be written as
            \begin{widetext}
                \begin{align}\label{eq:connection_relation_approx_1}
                    \begin{split}
                        \hat{A}_l^{\left(1,\mathrm{out}\right)} &= 2 \I \sum_{n,k} \ocg_{ln}  \sqrt{\Lambda^{\left(2\right)}_{n}} 
                            \sqrt{\Lambda^{\left(\mathrm{SU}\right)}_{k}} \Im\of{\och_{nk}} \left[\hat{A}_k^{\left(\mathrm{SU},\mathrm{in}\right)}  + \left( \hat{A}_k ^{\left(\mathrm{SU},\mathrm{in}\right)}\right)^{\dagger} \right] \\
                            &\qquad+ \frac{1}{2}\sum_{n,k} \ocg_{ln} \sqrt{\frac{\Lambda^{\left(2\right)}_{n}}{\Lambda^{\left(\mathrm{SU}\right)}_{k}}}
                            \left[  \och_{nk} \hat{A}_k^{\left(\mathrm{SU},\mathrm{in}\right)}  - \och_{nk}^* \left( \hat{A}_k ^{\left(\mathrm{SU},\mathrm{in}\right)}\right)^{\dagger}\right] \\
                            &\qquad+ \frac{1}{2}\sum_{n,k} \ocg_{ln} \sqrt{\frac{\Lambda^{\left(\mathrm{SU}\right)}_{k}}{\Lambda^{\left(2\right)}_{n}}} \och_{nk}
                             \left[\hat{A}_k^{\left(\mathrm{SU},\mathrm{in}\right)}  + 
                                \left( \hat{A}_k ^{\left(\mathrm{SU},\mathrm{in}\right)}\right)^{\dagger}\right] \\
                         &\qquad+ \frac{1}{4} \sum_{n,k} \ocg_{ln} \och_{nk}  \frac{1}{ \sqrt{\Lambda^{\left(2\right)}_{n}}}  \frac{1}{\sqrt{\Lambda^{\left(\mathrm{SU}\right)}_{k}}} \hat{A}_k^{\left(\mathrm{SU},\mathrm{in}\right)}.
                    \end{split}
                 \end{align}
            \end{widetext}
            Note that in the expression above, we have applied the
            Bogoliubov transformations for the output operators
            of the entire interferometer
            [compare \Cref{eq:bogoliubov_plain}]
            in order to express
            the output Schmidt operators of the first
            crystal $ \hat{A}_l^{\left(1,\mathrm{out}\right)} $ in terms
            of the input Schmidt operators of the
            entire interferometer $\hat{A}_k^{\left(\mathrm{SU},\mathrm{in}\right)}$.

            \par The conditions of the large number of photons in the mode 
            (large eigenvalues) $\Lambda^{\left(2\right)}_n \gg 1$, 
            $\Lambda^{\left(\mathrm{SU}\right)}_n \gg 1$ 
            imply that the last term in
            \Cref{eq:connection_relation_approx_1} is small and can be neglected.
            The first term of the \Cref{eq:connection_relation_approx_1} is
            proportional to the imaginary part of overlap coefficients
            $\Im\of{\och_{nk}}$.
            If focusing is done carefully, see \Cref{sec:gh_delta_z_phase},
            this imaginary part
            approaches zero [or, equivalently, the phase
            $\arg\of{\och_{nk}}$ approaches $0$ or $\pi$],
            so that the first term can be omitted
            (compare also \Cref{fig:omatsplots}).
            After neglecting these two terms, the
            remaining expression for the Schmidt mode operator reads
            \begin{widetext}
                \begin{align}\label{eq:connection_relation_approx_2}
                        \hat{A}_l^{\left(1,\mathrm{out}\right)} &= 
                            \frac{1}{2}\sum_{n,k} \ocg_{ln} \left\lbrace\sqrt{\frac{\Lambda^{\left(2\right)}_{n}}{\Lambda^{\left(\mathrm{SU}\right)}_{k}}}
                            \left[  \och_{nk} \hat{A}_k^{\left(\mathrm{SU},\mathrm{in}\right)}  - \och_{nk}^* \left( \hat{A}_k ^{\left(\mathrm{SU},\mathrm{in}\right)}\right)^{\dagger}\right]
                            +  \sqrt{\frac{\Lambda^{\left(\mathrm{SU}\right)}_{k}}{\Lambda^{\left(2\right)}_{n}}} \och_{nk}
                             \left[\hat{A}_k^{\left(\mathrm{SU},\mathrm{in}\right)}  +
                                \left( \hat{A}_k ^{\left(\mathrm{SU},\mathrm{in}\right)}\right)^{\dagger}\right]\right\rbrace.
                 \end{align}
            \end{widetext}
            These two remaining terms only contain the \emph{relative} 
            mean photon numbers of the modes
            of the amplifier and the interferometer.
            This means that in order to apply the discussed theoretical
            approach experimentally, it is sufficient to know the
            intensity in each Schmidt mode of the amplifier 
            $I_n^{\left(2\right)}$ and the
            interferometer $I_k^{\left(\mathrm{SU}\right)}$ since their 
            ratio follows the photon
            number (eigenvalue) ratio:
            $I_n^{\left(2\right)}/I_k^{\left(\mathrm{SU}\right)}
            =\Lambda_n^{\left(2\right)}/\Lambda_k^{\left(\mathrm{SU}\right)}$.

            \par Furthermore, in the case of the diagonal $\och$-matrix,
            \begin{align}\label{eq:h_eq_delta}
                \och_{nk}=\pm\delta_{nk},
            \end{align}
            where $\delta_{nk}$ is the Kronecker delta, 
            \Cref{eq:connection_relation_approx_2} simplifies to 
            \begin{widetext}
                \begin{align}\label{eq:A_approx_no_h}
                    \hat{A}_l^{\left(1,\mathrm{out}\right)} &= 
                        \frac{1}{2}\sum_{n} \pm\ocg_{ln} \left\lbrace\sqrt{\frac{\Lambda^{\left(2\right)}_{n}}{\Lambda^{\left(\mathrm{SU}\right)}_{n}}}
                        \left[  \hat{A}_n^{\left(\mathrm{SU},\mathrm{in}\right)} - \left( \hat{A}_n ^{\left(\mathrm{SU},\mathrm{in}\right)}\right)^{\dagger}\right]
                        +  \sqrt{\frac{\Lambda^{\left(\mathrm{SU}\right)}_{n}}{\Lambda^{\left(2\right)}_{n}}}
                         \left[\hat{A}_n^{\left(\mathrm{SU},\mathrm{in}\right)}  + 
                            \left( \hat{A}_n ^{\left(\mathrm{SU},\mathrm{in}\right)}\right)^{\dagger}\right]\right\rbrace.
                 \end{align}
            \end{widetext}
            Note that the $\pm$ in \Cref{eq:h_eq_delta} accounts for
            the fact that the sign
            of the modes is not well-defined, see 
            \Cref{sec:uniq_sdecomp}. This leads to random
            signs for the overlap coefficients $\och_{nk}$.
            In \Cref{eq:connection_relation_approx_2} this randomness does 
            not pose an issue since the overlap coefficients are 
            multiplied with the corresponding Schmidt-mode operators
            which contain the mode functions from which the random sign
            originates. This sign must however be taken into account for
            the approximation expressed by \Cref{eq:h_eq_delta}.
            Regardless, as will be seen below, the arbitrary sign does not
            affect the results for the quadrature variance.            
            Physically, the situation under which the approximation
            as written in 
            \Cref{eq:h_eq_delta} holds occurs when the output Schmidt modes
            of the second crystal 
            and of the interferometer are sufficiently similar to each other.

            \par The generic quadrature operator $\hat{P}_{\vartheta,l}$ 
            can be obtained by plugging \Cref{eq:A_approx_no_h} into 
            \Cref{eq:gen_quad_def} and simplifying the result:
            \begin{widetext}
                \begin{align}
                    \hat{P}_{\vartheta,l} &= \sum_n \left[\Re\of{\ocg_{ln}\E^{-\I\frac{\vartheta}{2}}}
                        \sqrt{\frac{\Lambda_n^{\left(\mathrm{SU}\right)}}{\Lambda_n^{\left(2\right)}}} 
                        \hat{Y}_n^{\left(\mathrm{SU},\mathrm{in}\right)} - \Im\of{\ocg_{ln}\E^{-\I\frac{\vartheta}{2}}}
                        \sqrt{\frac{\Lambda_n^{\left(2\right)}}{\Lambda_n^{\left(\mathrm{SU}\right)}}} 
                        \hat{X}_n^{\left(\mathrm{SU},\mathrm{in}\right)}\right],
                \end{align}
            with $\hat{X}_n^{\left(\mathrm{SU},\mathrm{in}\right)}$
            and $\hat{Y}_n^{\left(\mathrm{SU},\mathrm{in}\right)}$
            as defined in \Cref{eq:X_op_A,eq:Y_op_A} in terms of
            the input Schmidt operators
            of the entire interferometer.
            Clearly, $\langle\hat{P}_{\vartheta,l}\rangle=0$ and
            \begin{align}\label{eq:Psquared_general}
                \langle\hat{P}_{\vartheta,l}^2\rangle
                    &= \sum_n \abs{\nophase{\ocg}_{ln}}^2 \left[
                        \cos^2\of{\arg\of{\nophase{\ocg}_{ln}}+\frac{\phi-\vartheta}{2}}
                    \frac{\Lambda_n^{\left(\mathrm{SU}\right)}}{\Lambda_n^{\left(2\right)}}
                    + \sin^2\of{\arg\of{\nophase{\ocg}_{ln}}+\frac{\phi-\vartheta}{2}}
                    \frac{\Lambda_n^{\left(2\right)}}{\Lambda_n^{\left(\mathrm{SU}\right)}}\right],
            \end{align}
            \end{widetext}
            where the known phase dependence of the overlap
            matrix elements $\ocg_{ln}$ as written in \Cref{eq:Psquared_general}
            has been applied so that $\nophase{\ocg}_{ln}$ no
            longer depends on $\phi$:
            \begin{align}
                \ocg_{ln}=\nophase{\ocg}_{ln} \E^{\I\frac{\phi}{2}},
            \end{align}
            compare \Cref{eq:g_with_phase_imperfect,eq:nophase_notation}.
            It should also be noted that the $\Lambda_n^{\left(\mathrm{SU}\right)}$
            in \Cref{eq:Psquared_general}
            also depend on $\phi$.
            In principle, \Cref{eq:Psquared_general} allows for the reconstruction of
            the quadrature variance
            $\Dvar\hat{P}_{\vartheta,l}=\langle\hat{P}_{\vartheta,l}^2\rangle$
            for any quadrature angle $\vartheta$, given the overlap matrix
            elements $\nophase{\ocg}_{ln}$, the eigenvalues
            of the amplifier $\Lambda_n^{\left(\mathrm{2}\right)}$
            and the eigenvalues of the interferometer
            $\Lambda_n^{\left(\mathrm{SU}\right)}$ at any known phase $\phi$.
            Then, from $\Dvar\hat{P}_{\vartheta,l}$, the
            levels of squeezing and anti-squeezing 
            can be immediately obtained, see \Cref{eq:sq_def,eq:asq_def}.

            \par In general, for the unbalanced interferometer, $\nophase{\ocg}_{ln}=\int\!\dd q
            \left[u^{\left(1\right)}_k\of{q}\right]^{*}
            \nophase{\psi}^{\left(2\right)}_m\of{q}$ is still a complex number, even after
            the $\phi$-dependence has been split off.
            However, if the focusing element is moved carefully by some distance
            $\delta z$  away from the optimal distance in the gain-balanced case,
            see \Cref{sec:unbalanced_interf}, it is possible to
            reach $\nophase{\ocg}_{ln}\approx \pm 1$ for several low-order modes for the
            gain-unbalanced interferometer.
            For the present example, this can be seen directly from \Cref{fig:omatsplots},
            where for the bright fringe ($\phi\approx 0$),
            $\ocg_{ln}=\nophase{\ocg}_{ln}\E^{\I\phi/2}\approx \pm 1$,
            while for the dark fringe ($\phi\approx\pi$),
            $\nophase{\ocg}_{ln}\E^{\I\phi/2}\approx \pm \I$, in both cases
            for low-order modes. In both cases, this implies
            $\nophase{\ocg}_{ln}\approx \pm 1$ 
            and therefore 
            $\arg\of{\nophase{\ocg}_{ln}}\approx m\pi$, for $m\in\mathbb{Z}$.
            Since both $\cos^2$ and $\sin^2$ are $\pi$-periodic, $\arg\of{\nophase{\ocg}_{ln}}$
            may therefore be neglected in \Cref{eq:Psquared_general}, if only the low-order
            modes have significant contributions to $\langle\hat{P}_{\vartheta,l}^2\rangle$.
            For a more detailed analysis
            of the behavior of the phases of the overlap coefficients
            in dependence on $\delta z$ see \Cref{sec:gh_delta_z_phase}. 

            \par Finally, neglecting the term $\arg\of{\nophase{\ocg}_{ln}}$ and
            considering the dark fringe with the phase $\phi_{\mathrm{df}}\approx\pi$, 
            the second contribution in \Cref{eq:Psquared_general}
            vanishes for $\vartheta=\phi_{\mathrm{df}}/2+\pi n$, where
            $n\in\mathbb{Z}$, while the first contribution vanishes for
            $\vartheta=\phi_{\mathrm{df}}/2+\pi \left(n+1/2\right)$.
            At the dark fringe, it would be expected that
            $\Lambda_n^{\left(2\right)}>\Lambda_n^{\left(\mathrm{SU}\right)}$,
            compare \Cref{fig:eigvals},
            which means that the maximal levels of squeezing and anti-squeezing 
            can be obtained as [compare \Cref{eq:sq_def,eq:asq_def}]:
            \begin{subequations}
                \begin{align}
                    \Sq_{\mathrm{df},l} &= 10\log_{10}\ofb{\sum_n \abs{\ocg_{ln}}^2 
                        \frac{\Lambda_n^{\left(\mathrm{SU,df}\right)}}{\Lambda_n^{\left(2\right)}}}\unit{\deci\bel}, 
                        \label{eq:hg_Sqdf} \\
                    \ASq_{\mathrm{df},l} &= 10\log_{10}\ofb{\sum_n \abs{\ocg_{ln}}^2 
                        \frac{\Lambda_n^{\left(2\right)}}{\Lambda_n^{\left(\mathrm{SU,df}\right)}}}\unit{\deci\bel},
                        \label{eq:hg_ASqdf}
                \end{align}
            \end{subequations}
            Similarly, at the bright fringe with phase $\phi_{\mathrm{bf}}\approx 0$,
            the roles of the eigenvalues are reversed and it should be expected
            that $\Lambda_n^{\left(2\right)}<\Lambda_n^{\left(\mathrm{SU}\right)}$.
            Hence, the maximal levels of squeezing and anti-squeezing are given by
            \begin{subequations}
                \begin{align}
                    \Sq_{\mathrm{bf},l} &= 10\log_{10}\ofb{\sum_n \abs{\ocg_{ln}}^2 
                        \frac{\Lambda_n^{\left(2\right)}}{\Lambda_n^{\left(\mathrm{SU,bf}\right)}}}\unit{\deci\bel}, 
                        \label{eq:hg_Sqbf}\\
                    \ASq_{\mathrm{bf},l} &= 10\log_{10}\ofb{\sum_n \abs{\ocg_{ln}}^2 
                        \frac{\Lambda_n^{\left(\mathrm{SU,bf}\right)}}{\Lambda_n^{\left(2\right)}}}\unit{\deci\bel}.
                        \label{eq:hg_ASqbf}
                \end{align}
            \end{subequations}
            It should be noted that \Crefrange{eq:hg_Sqdf}{eq:hg_ASqbf}
            are estimations based on the assumptions that the
            respective orderings
            $\Lambda_n^{\left(2\right)}>\Lambda_n^{\left(\mathrm{SU}\right)}$
            and
            $\Lambda_n^{\left(2\right)}<\Lambda_n^{\left(\mathrm{SU}\right)}$
            of the eigenvalues in the bright and dark fringe, respectively,  hold
            for sufficiently large $n$, while the  $ \och_{nk}$ matrix is diagonal.
            In general,
            it is otherwise not possible to obtain
            simple expressions for the levels of squeezing and
            anti-squeezing, and instead, 
            \Cref{eq:min_P_theta_res,eq:max_P_theta_res,eq:sq_def,eq:asq_def} 
            together with \Cref{eq:exact_reco_AA} and
            \Cref{eq:exact_reco_AdA}
            have to be used directly 
            to obtain expressions for $\Sq_l$
            and $\Asq_l$.

        \subsection{Comparison of the different approaches}
            Analytically, the level of squeezing and
            anti-squeezing of the output quadratures
            of the first crystal can be directly obtained from the 
            Schmidt decomposition and the
            resulting Bogoliubov transformations of the first crystal.
            Plugging \Cref{eq:bogoliubov_plain} into
            \Cref{eq:genPquadvar} yields
            \begin{align}\label{eq:dvar_quadp_c1}
                 \Dvar \hat{P}_{\vartheta,l} 
                    = 1 + 2\Lambda_l^{\left(1\right)}
                    + 2 \sqrt{\tilde{\Lambda}_l^{\left(1\right)} 
                    \Lambda_l^{\left(1\right)}} \cos\of{\vartheta},
            \end{align}
            while the vacuum quadratures variances are again
            $\Dvar\hat{P}_{\vartheta,l}^{\left(\mathrm{vac}\right)}=1$.
            Using the fact that the eigenvalues can be parameterized as
            $\Lambda_{l}=\sinh^2\of{r_l}$ and
            $\tilde{\Lambda}_{l}=\cosh^2\of{r_l}$~\cite{PRR5,Christ_2013},
            where $r_l$ is associated with the gain in 
            mode $l$ according to Schmidt-mode theory~\cite{PRA91,APL2025},
            \Cref{eq:dvar_quadp_c1} may be rewritten as
            \begin{align}
                 \Dvar \hat{P}_{\vartheta,l} 
                    = \cosh\of{2r_l}+\sinh\of{2r_l}\cos\of{\theta}.
            \end{align}
            The levels of squeezing and anti-squeezing are therefore:
            \begin{subequations}
                \begin{align}
                    \Sq_l &= -\frac{20 r_l}{\ln\of{10}}  \unit{\deci\bel}, \label{eq:sq_direct_c1} \\
                    \ASq_l &= \frac{20 r_l}{\ln\of{10}} \unit{\deci\bel}. \label{eq:asq_direct_c1}
                \end{align}
            \end{subequations}
            Note that $20/\ln\of{10}\approx 8.686$.

            \par
\begin{figure*}[th!]%
                \centering%
                \subfloat{\includegraphics[width=0.5\linewidth]{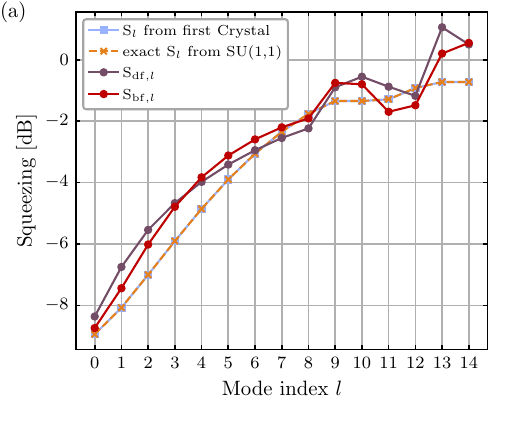}}%
                \subfloat{\includegraphics[width=0.5\linewidth]{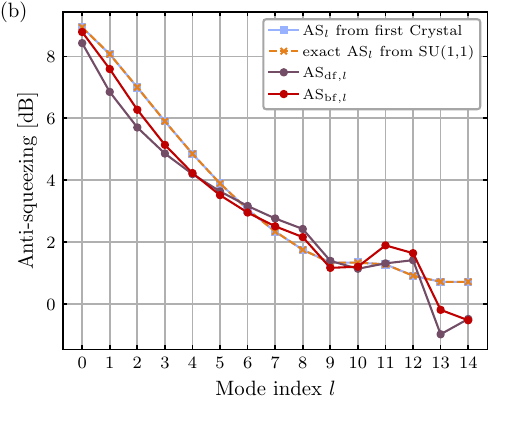}}%
                \caption{(a)~Comparison of the level of squeezing 
                    in each Schmidt mode $l$ of the first crystal (squeezer)
                    obtained directly at the
                    output of the first crystal using \Cref{eq:sq_direct_c1}, extracted from the
                    output of the SU(1,1) interferometer using the exact approach described in
                    \Cref{sec:exact_reconstruction} and using the high-gain approximation
                    using \Cref{eq:hg_Sqdf,eq:hg_Sqbf}. As expected, the exact procedure coincides
                    with the squeezing obtained directly using the first crystal eigenvalues.
                    The high-gain approximation results in underestimated values for the squeezing.
                    (b)~Comparison of the level of anti-squeezing obtained directly at the
                    output of the first crystal using \Cref{eq:asq_direct_c1}, extracted from the
                    output of the SU(1,1) interferometer using the exact approach described in
                    \Cref{sec:exact_reconstruction} and using the high-gain approximation
                    using \Cref{eq:hg_ASqdf,eq:hg_ASqbf}. As was the case for the
                    level of squeezing,
                    the exact procedure coincides with the 
                    anti-squeezing obtained directly using the first
                    crystal eigenvalues, while the high-gain approximation results in underestimated values.
                }%
                \label{fig:squeezingplots}%
            \end{figure*}

 In \Cref{fig:squeezingplots}, the levels of squeezing and anti-squeezing 
            obtained from the methods described in
            \Cref{sec:exact_reconstruction,sec:high_gain_approx_reconstruction}
            are compared against the exact values evaluated using
            \Cref{eq:sq_direct_c1,eq:asq_direct_c1}.
            As expected, the levels of squeezing and anti-squeezing obtained
            using the exact method presented in \Cref{sec:exact_reconstruction}
            coincide with the values obtained from 
            \Cref{eq:sq_direct_c1,eq:asq_direct_c1}.
            Contrary to that, the levels of squeezing and
            anti-squeezing obtained using the high-gain approximation as
            presented in \Cref{sec:high_gain_approx_reconstruction}
            underestimate the expected values for almost all shown mode
            indices.
            For the level of squeezing, this means that the high-gain approximation 
            results in larger \revA{(``less negative'')} values for $\Sq_n$ compared to the
            theoretically exact values,
            while for the anti-squeezing, the obtained $\ASq_n$ values 
            are smaller.
            \revA{A brief discussion of the extension of the presented 
            formalism to include losses is presented in \Cref{sec:lossinclusion}.}
             
    \section{CONCLUSION}
        We have presented a rigorous theoretical analysis of the spatial Schmidt mode
        structure of an unbalanced multimode SU(1,1)
        interferometer. We analyzed the Schmidt mode shapes 
        for each crystal separately and for the entire interferometer
        and demonstrated a strong dissimilarity in the phase profile
        of the input and output Schmidt modes with increasing gain.  
        Considering the complex interplay between the input and output modes
        of a cascaded crystal setup, we extended the phase-sensitive amplification
        approach to the multimode scenario and presented a generalized theoretical
        framework for measuring multimode squeezing based on the direct 
        intensity measurement.  
        To make our framework experimentally feasible, we considered
        the high-gain amplifier approximation, which has been experimentally 
        realized in~\refcite{multimodesqueezing} and found to be in good agreement
        with theoretical predictions.
        In addition, for the unbalanced interferometer, we discussed experimental 
        optimizations and showed how shifting the focal point of the focusing
        element affects the overlap coefficients and
        helps to improve the visibility.

        \par Our work provides a theoretical background for multimode PDC setups
        and the presented theory can be extended for a wide range of
        experiments using multimode PDC light. 
        \revA{For example, in addition to multimode squeezing measurements,
        the presented framework can be used to improve the phase sensitivity 
        in quantum-enhanced metrology by eliminating mode mismatch
        (which acts as internal losses and reduces phase sensitivity).}
        In addition, the lack of experiments on the detailed study 
        of the phases of the Schmidt-modes and the overlap coefficients 
        can be filled and motivated by the current work.  
        \revA{Indeed, the moduli of the output Schmidt modes can be found 
        by decomposing the covariance matrix \cite{PhysRevA.95.053814,Chekhova2025}. 
        However, the measurement of the Schmidt-mode phases requires
        additional investigations. Moreover, since the input and output
        Schmidt modes are generally different, their detailed experimental 
        characterization is an important and still open question.}
        
        \par Finally, a different set of modes may be more
        beneficial and simplify the measurement procedure dramatically:
        For example, using the full two-dimensional setup gives access
        to the azimuthal degree of freedom so that the Schmidt modes will be
        given by modes of the form $u_{mn}\of{q}\E^{\I n\phi}$~\cite{PRA91,APL2025}, where
        $q$ now refers to the radial component of the transverse wave-vector, while
        $\phi$ is the azimuthal angle. Clearly, these modes carry an orbital
        angular momentum of $n$, and therefore, if the measurement basis is
        extended to orbital angular momentum modes, the cross-coupling 
        between the modes inside the interferometer is reduced to being between modes
        with different radial indices $m$, which then however have the 
        same orbital angular momentum $n$. 
        \revA{However, it should be noted that this treatment would
        strongly increase the numerical complexity compared to the case
        where only one transverse dimension is considered.}

    \section*{ACKNOWLEDGMENTS}
        We would like to thank M.~Chekhova, I.~Barakat, and
        M.~Kalash for insightful discussions. We acknowledge
        financial support of the Deutsche Forschungsgemeinschaft
        (DFG) via Project SH 1228/3-1 and via the TRR 142/3
        (Project No.\@ 231447078, Subproject No.\@ C10) and we
        thank the PC2 (Paderborn Center for Parallel Computing)
        for providing computation time. We acknowledge support for 
        the publication costs by the Open Access Publication Fund 
        of Paderborn University.

    \section*{DATA AVAILABILITY}
        The data used to generate the figures in this article are available 
        from \refcite{datazenodo}.

    \appendix

    \section{THE JOINT SCHMIDT DECOMPOSITION}%
            \label[appendix]{sec:jsdecomp}

        \subsection{Uniqueness of the Schmidt modes}\label{sec:uniq_sdecomp}
            Compared to a simple Schmidt decomposition of only one of the
            transfer functions, for example only of $\beta$, as written in
            \Cref{eq:jointsdecompbeta} on its own, the joint decomposition as written
            in \Cref{eq:jointsdecompbeta,eq:jointsdecompeta} imposes
            restrictions on the sets $\set{\psi_n}$ and $\set{u_n}$ of possible modes.
            To investigate this, it is helpful to draw analogies between
            the equations written in the main text for the continuous case over $q$
            and their discretized versions.
            In the discretized case, the Schmidt decomposition corresponds to the
            singular value decomposition (SVD).
            SVD algorithms also form the base for the numerical
            computation of the joint decomposition, see \Cref{sec:numremarks_jsdecomp}
            below.
            The transfer functions $\tilde{\eta}$ and $\beta$
            as defined in \Cref{eq:as_dagger_sol,eq:ai_dagger_sol}, are then
            represented by complex-valued matrices.
            More precisely, if we
            label the matrices corresponding to $\tilde{\eta}$ and $\beta$
            by their corresponding uppercase letters, $\tilde{\Eta}$ and
            $\Beta$, respectively, and if $n$ discretization
            points $q_1,\ldots,q_{n}$ are used for the underlying 
            $q$-lattices, one finds that
            $\tilde{\Eta},\Beta\in\mathbb{C}^{n\times n}$ and that the joint
            Schmidt decomposition
            reads~\cite{PhysRevA.93.062115,PhysRevA.94.062109,Kopylov2025}:
            \begin{subequations}
                \begin{align}
                    \Beta &= U \Sigma \Psi^{\T}, \label{eq:jsdecompBeta} \\
                    \tilde{\Eta} &= U \tilde{\Sigma} \Psi^{\H}, \label{eq:jsdecompEtatilde} 
                \end{align}
            \end{subequations}
            where $U$ and $\Psi$ are unitary matrices and we denote the
            transpose and hermitian conjugates of matrices with~$^{\T}$
            and~$^{\H}$, respectively.
            The columns $\vec{u}$ of $U$ and the columns $\vec{\psi}$ of
            $\Psi$ correspond to the discretized versions
            of the mode functions $u\of{q}$ and $\psi\of{q'}$, respectively:
            \begin{subequations}
                \begin{align}
                    U &= \begin{bmatrix}
                        \vec{u}_1 & \vec{u}_2 & \cdots & \vec{u}_{n}
                    \end{bmatrix}, \\
                    \Psi &= \begin{bmatrix}
                        \vec{\psi}_1 & \vec{\psi}_2 & \cdots & \vec{\psi}_{n}
                    \end{bmatrix},
                \end{align}
            \end{subequations}
            so that
            \begin{subequations}
                \begin{align}
                    u_j\of{q} \leftrightarrow \vec{u}_j &= \begin{bmatrix}
                        u_j\of{q_1}\\
                        u_j\of{q_2}\\
                        \vdots\\
                        u_j\of{q_{n}}\\
                    \end{bmatrix}, \\
                    \psi_j\of{q} \leftrightarrow \vec{\psi}_j &= \begin{bmatrix}
                        \psi_j\of{q_1}\\
                        \psi_j\of{q_2}\\
                        \vdots\\
                        \psi_j\of{q_{n}}\\
                    \end{bmatrix}.
                \end{align}
            \end{subequations}
            Furthermore, 
            \begin{subequations}
                \begin{align}
                    \Sigma=\diag\of{\sqrt{\Lambda_1}, \sqrt{\Lambda_2}, \ldots, \sqrt{\Lambda_n}}, \label{eq:sigma_sings}\\
                    \tilde{\Sigma}=\diag\of{\sqrt{\tilde{\Lambda}_1}, \sqrt{\tilde{\Lambda}_2}, \ldots, 
                        \sqrt{\tilde{\Lambda}_n}} \label{eq:sigma_sings_tilde}
                \end{align}
            \end{subequations}
            are diagonal matrices with positive
            entries containing the singular values of $\Beta$ and
            $\tilde{\Eta}$, respectively. Note that we neglect additional
            scaling factors such as the lattices spacing $\dd q$ which appears
            due to the integrals when making these analogies for simplicity.
            
            \par For non-degenerate eigenvalues $\Lambda_n$ and 
            $\tilde{\Lambda}_n$, it is well-known that the 
            vectors $\vec{u}_n$ and $\vec{\psi}_n$ resulting from the singular 
            value decomposition are uniquely defined up to a phase 
            factor~\cite{trefethen1997}.
            In the case of degenerate eigenvalues,
            the modes have a higher degree of freedom with respect to the phase
            rotation and it is possible 
            to perform unitary rotations in the corresponding subspace of the 
            degenerate modes. For the rest of this section, we will focus only on the
            discussion of the case of non-degenerate eigenvalues since for
            the typical parameters discussed in the main text of this work,
            the low-order eigenvalues which have the highest contributions
            are usually non-degenerate, compare also \Cref{fig:eigvals}.
            Furthermore, note that due to the connection between the
            eigenvalues written in \Cref{eq:connection_L_Ltilde},
            if the $\Lambda_n$ are all non-degenerate, so are all the
            $\tilde{\Lambda}_n$ as well, and vice versa\footnote{This may not be
            necessarily true numerically due to floating-point errors. However,
            these usually only become relevant for extremely large eigenvalues
            $\Lambda_n\ggg1$. As such, this issue can, in principle, be
            mitigated by appropriately re-scaling $\beta$ and $\tilde{\eta}$.}.
            Similarly, if $\Lambda_{j}$ is an $m$-fold degenerate eigenvalue,
            $\tilde{\Lambda}_{j}$ must also be $m$-fold degenerate.
            
            \par Transferring this to the case of continuous $q$ variables,
            we will assume in the
            following that the modes $u_n$ and $\psi_n$ are also defined up
            to a constant phase.
            The joint Schmidt decomposition of the two transfer functions results
            in a pair $P$ of two sets of modes $\set{u_n}$ and $\set{\psi_n}$:
            $P=\left(\set{u_n}, \set{\psi_n}\right)$.
            Then, given another independent pair of sets of input and output modes,
            $P'=\left(\set{u_n'}, \set{\psi_n'}\right)$,
            the joint Schmidt decomposition for the former sets $P$ takes the form
            as written in \Cref{eq:jointsdecompbeta,eq:jointsdecompeta}, while
            for $P'$, the mode functions therein would be replaced with 
            $u_n'$ and $\psi_n'$. However, due to the fact that the mode functions
            are only defined up to some phase, the modes in $P'$ may be expressed
            in terms of the modes of $P$ and they must be connected via
            $u_n'\of{q}=u_n\of{q}\E^{\I\xi_{u_n}}$ and
            $\psi_n'\of{q}=\psi_n\of{q}\E^{\I\xi_{\psi_n}}$.
            Thus, the decomposition for $P'$ may be written as:
            \begin{subequations}
                \begin{align}
                    \beta\of{q,q'} &= \sum_{n} \sqrt{\Lambda_{n}} u_{n}\of{q}\psi_{n}\of{q'}
                        \E^{\I\xi_{u_n}+\I\xi_{\psi_n}}, \label{eq:jointsdecompbeta_second} \\
                    \tilde{\eta}\of{q,q'} &= \sum_{n} \sqrt{\tilde{\Lambda}_{n}}
                        u_{n}\of{q}\psi_{n}^{*}\of{q'}
                        \E^{\I\xi_{u_n}-\I\xi_{\psi_n}}. \label{eq:jointsdecompeta_second}
                \end{align}
            \end{subequations}
            Multiplying both transfer functions with $u_m\of{q}$ and $\psi_m\of{q'}$,
            integrating over the $q$-variables and then comparing the result 
            obtained for 
            \Cref{eq:jointsdecompbeta,eq:jointsdecompeta}
            with that obtained for
            \Cref{eq:jointsdecompbeta_second,eq:jointsdecompeta_second}
            yields a system of coupled equations for the possible values
            of $\xi_{\psi_m}$ and $\xi_{\psi_m}$ which can be written as
            \begin{align}
                \begin{cases}
                    \begin{aligned}
                        \E^{\I\xi_{u_m}+\I\xi_{\psi_m}} &= 1 \\
                        \E^{\I\xi_{u_m}-\I\xi_{\psi_m}} &= 1.
                    \end{aligned}
                \end{cases}
            \end{align}
            The solutions of this set of equations are of the form
            \begin{align}
                \begin{cases}
                    \begin{aligned}
                        \xi_{\psi_m} &= m\pi \\
                        \xi_{u_m} &= \xi_{\psi_m} + 2l\pi,
                    \end{aligned}
                \end{cases}
            \end{align}
            for $l,m\in\mathbb{Z}$.
            Thus, in conclusion, the modes of $P$ and $P'$ only differ
            in the sense that $u_n$ and $\psi_n$ may simultaneously
            (but for each $n$ independently)
            be replaced by $-u_n$ and $-\psi_n$, respectively.
            The modes are therefore actually only unique up to their sign,
            with the additional restriction that the signs of
            both the input modes and the output modes must be changed
            together to preserve the decomposition. In more direct terms,
            for each $n$, a numerical procedure may result either in 
            $u_n$ and $\psi_n$ or in the same pair with the opposite
            signs: $-u_n$ and $-\psi_n$.

        \subsection{Remarks on the numerical computation}\label{sec:numremarks_jsdecomp}
            Equations~(\ref{eq:jsdecompBeta}) and~(\ref{eq:jsdecompEtatilde})
            describe a special joint \emph{singular value decomposition}
            (SVD) of the
            matrices $\Beta$ and $\tilde{\Eta}$.
            Methods for finding such a decomposition
            have for example been described in
            \refcite{PhysRevA.93.062115,Kopylov2025,Houde2024}.
            Note that generally, it is not possible to simply perform two
            independent singular value decompositions of $\Beta$ and $\tilde{\Eta}$, 
            since these are usually not unique and the resulting
            matrices $U$ and $\Psi$ will generally not fulfill
            \Cref{eq:jsdecompBeta,eq:jsdecompEtatilde}
            above~\cite{PhysRevA.94.062109}.
            Additionally, care must be taken when applying numerical methods
            due to degeneracies which may appear in the singular value
            matrices $\Sigma$ and $\tilde{\Sigma}$ 
            [see \Cref{eq:jsdecompBeta,eq:jsdecompEtatilde}]
            and because numerical methods may not necessarily preserve
            symmetries present in their input matrices.

            \par One way of obtaining the joint decomposition
            as written in \Cref{eq:jsdecompBeta,eq:jsdecompEtatilde} is
            as follows. First, two independent SVDs of $\Beta$ and $\tilde{\Eta}$
            are performed\footnote{Note that the notation used here is
            slightly different than in
            \refcite{PhysRevA.93.062115,Kopylov2025,PhysRevA.94.062109}.
            When comparing with these works, note that ${\tilde{\Eta}\leftrightarrow E}$
            and ${\Beta\leftrightarrow F}$.
            }\textsuperscript{,}\footnote{%
            Note that numerical implementations of the SVD may return either
            $V_0^{\H}$ and $\tilde{V}^{\H}$ or 
            $V_0$ and $\tilde{V}$
            as the right-hand side unitary.}:
            \begin{subequations}
                \begin{align}
                    \Beta &= \UU \Sigma V_0^{\H},
                        \label{eq:jsdecompBeta_first} \\
                    \tilde{\Eta} &= \tilde{\UU} \tilde{\Sigma}
                        \tilde{V}^{\H},
                        \label{eq:jsdecompEtatilde_first} 
                \end{align}
            \end{subequations}
            yielding the unitary matrices $\UU$, $\tilde{\UU}$
            (with, in general,  $\UU\neq\tilde{\UU}$),
            $V_0$ and $\tilde{V}$,
            as well the diagonal matrices $\Sigma$ and
            $\tilde{\Sigma}$ which contain the singular values, see
            \Cref{eq:sigma_sings,eq:sigma_sings_tilde}.
            Then, the matrix
            \begin{align}\label{eq:def_K}
                K = \UU^{\H} \tilde{\UU}
            \end{align}
            and its hermitian conjugate $K^{\H}$
            are unitary, block-diagonal and commute with $\Sigma$
            and $\tilde{\Sigma}$.
            Expressing the $n\times n$ identity matrix $I_n$ as $I_n = K K^{\H}$
            and inserting it into \Cref{eq:jsdecompBeta_first} yields
            $\Beta = \UU I_n \Sigma V_0^{\H}
            =\left(\UU K \right) \Sigma  \left[K^{\H}
            V_0^{\H}\right]$,
            which brings the SVDs in
            \Cref{eq:jsdecompBeta_first,eq:jsdecompEtatilde_first}
            to the form
            \begin{subequations}
                \begin{align}
                    \Beta &= \tilde{\UU} \Sigma V^{\H}, \label{eq:jdecomp_form_b} \\
                    \tilde{\Eta} &= \tilde{\UU} \tilde{\Sigma}
                            \tilde{V}^{\H}, \label{eq:jdecomp_form_h}
                \end{align}
            \end{subequations}
            where $V= V_0 K$.
            Equations~(\ref{eq:jdecomp_form_b}) and~(\ref{eq:jdecomp_form_h})
            describe a special SVD of the matrices $\Beta$ and $\tilde{\Eta}$,
            where the left-hand side unitary $\tilde{\UU}$ is identical for both matrices.
            This SVD can serve as the starting point for the algorithm
            described in \refcite{PhysRevA.93.062115} to bring
            \Cref{eq:jdecomp_form_b,eq:jdecomp_form_h} to the form of
            \Cref{eq:jsdecompBeta,eq:jsdecompEtatilde}.
            Alternatively, $K$ may be defined as $K=\tilde{\UU}^{\H}\UU$ 
            [instead of as written in \Cref{eq:def_K}] and inserted into
            \Cref{eq:jsdecompEtatilde_first} so that retracing the steps above, 
            but for $\tilde{\Eta}$, yields a decomposition of the same form as in 
            \Cref{eq:jdecomp_form_b,eq:jdecomp_form_h}.

            \par As another alternative, such a decomposition can be obtained
            as suggested in \refcite{PhysRevA.93.062115,Kopylov2025}, which is, by
            performing an eigendecomposition of $\Beta\Beta^{\H}$ or 
            $\tilde{\Eta}\tilde{\Eta}^{\H}$. According to \refcite{PhysRevA.93.062115},
            this yields $\Beta\Beta^{\H}=\UU_{1}\Sigma^2\UU_{1}^{\H}$
            or $\tilde{\Eta}\tilde{\Eta}^{\H}=\UU_{2}\tilde{\Sigma}^2\UU_{2}^{\H}$,
            where both $\UU_{1}$ and $\UU_{2}$ are candidates for
            the left-hand side unitary $\UU$
            in \Cref{eq:jdecomp_form_b,eq:jdecomp_form_h}.
            This is because
            $\tilde{\Eta}\tilde{\Eta}^{\H}$ and $\Beta\Beta^{\H}$ are connected
            via~\cite{PhysRevA.93.062115,Christ_2013,Kopylov2025}
            \begin{align}\label{eq:BBpIeqHH}
                \Beta\Beta^{\H}+I_n= \tilde{\Eta}\tilde{\Eta}^{\H},
            \end{align}
            meaning that any choice for $\UU_1$ or $\UU_2$ that diagonalizes 
            the matrix $\Beta\Beta^{\H}$
            or $\tilde{\Eta}\tilde{\Eta}^{\H}$, will also diagonalize
            the other matrix and is therefore a candidate for the left-hand side unitary
            of both SVDs.

            \par The next step is to compute the matrix~\cite{PhysRevA.93.062115}
            \begin{align}\label{eq:G_takagi}
                G = \tilde{V}^{\H} V^{*},
            \end{align}
            where $^{*}$ denotes the element-wise complex
            conjugate of a matrix. This matrix $G$ is unitary,
            symmetric and block-diagonal and commutes with $\Sigma$
            and $\tilde{\Sigma}$~\cite{PhysRevA.93.062115}.
            Such a matrix allows for a special SVD, the \emph{Takagi factorization},
            which here is of the form\footnote{Generally, the Takagi factorization
            of a symmetric matrix $G$ is defined as 
            $G=D S D^{\T}$~\cite{PhysRevA.93.062115,CHEBOTAREV2014380,Houde2024},
            where $D$ is unitary
            and $S$ is the diagonal matrix containing the
            singular values. However, for a unitary matrix, all singular values
            are equal to $1$, meaning $D=I_n$ and therefore, the Takagi factorization
            takes the form as in \Cref{eq:G_takagi}.}
            \begin{align}
                G = D D^{\T},
            \end{align}
            with a unitary matrix $D$, which here must be chosen to have
            the same block sizes as $G$. As suggested in \refcite{PhysRevA.93.062115},
            this can be done by computing the factorization $G_i=D_i D_i^{\T}$
            for each block $G_i$ of $G$
            separately and building $D$ from the blocks $D_i$. For the numerical
            computation, this also ensures that $D$ commutes with 
            $\Sigma$ and $\tilde{\Sigma}$.
            
            \par It was also suggested in 
            \refcite{PhysRevA.93.062115} 
            to perform the computation of $D$ as
            \begin{align}
                D=\left(V^{\H}\tilde{V}^{*}\right)^{\frac{1}{2}},
            \end{align}
            where $^{\frac{1}{2}}$ denotes the matrix square-root.
            Note that here, the matrix-square root also has to be applied
            for each block of the matrix
            $V^{\H}\tilde{V}^{*}$ separately,
            to preserve the block-diagonal structure of the 
            matrix, as suggested in \refcite{PhysRevA.93.062115}.
            It should however be noted that
            if the matrix square-root is computed \emph{numerically},
            it is not automatically guaranteed that the resulting square-root
            matrix is symmetric (meaning $D=D^{\T}$);
            but this is required for the proof of the decomposition, see 
            Eqs.~(A6)\crefrangeconjunction(A8) of \refcite{PhysRevA.93.062115}.
            Lastly, a numerical computation may also result in a matrix $D$
            which is not unitary,
            meaning that the resulting matrices $U$
            and $\Psi$ will also ultimately not be unitary, see 
            \Cref{eq:jsdecomp_U_UUD,eq:jsdecomp_Psi_DPsi} below\footnote{
                As an example for the abovementioned issues that may arise,
                consider the $2\times 2$ identity matrix 
                $I_2=\begin{bsmallmatrix}1 & 0 \\ 0 & 1 \end{bsmallmatrix}$, which is a unitary,
                symmetric and block-diagonal matrix consisting of two blocks 
                of size $1$. The matrix 
                $S=\begin{bsmallmatrix}1 & \I \\ 0 & -1 \end{bsmallmatrix}$
                is one of the matrix square-roots of $I_2$: $S^2=I_2$.
                However, $S$ is clearly neither symmetric, nor is it unitary,
                see for example
                $S^{\H}S=\begin{bsmallmatrix}1 & \I \\ -\I & 2 \end{bsmallmatrix}\neq I_2$.
                Furthermore, $S$ consist of a single block of size $2$.
                Generally, a numerical algorithm may therefore not necessarily
                preserve the symmetry, unitarity and/or the block-diagonal
                structure of the input matrix.
            }.

            \par Note that the computation of the Takagi factorization  
            also requires the computation of a matrix square-root, for example
            in the algorithm as described in \refcite{CHEBOTAREV2014380,PhysRevA.94.062109}.
            With this, the requirement that the symmetry of the input
            matrix is preserved when computing the matrix square root is lifted.
            The computation of the matrix square root
            can be for example performed using the eigendecomposition
            or using algorithms for the Schur decomposition (both 
            decompositions coincide for normal matrices).
            
            \par As noted in \refcite{CHEBOTAREV2014380}, the computation should
            also be performed for each block of the input matrix separately,
            as already described above.
            Let $Z$ be the block-diagonal unitary matrix
            whose matrix square-root $Z^{\frac{1}{2}}$ should be computed
            and let the blocks of $Z$ be labeled $Z_j$.
            The eigendecomposition (or Schur decomposition) of $Z_j$ is
            of the form $Z_j=K_j S_j K_j^{\H}$, with the unitary matrix
            $K_j$ and the diagonal eigenvalue matrix $S_j$, whose entries
            all have modulus $1$, meaning $S_j$ is itself unitary.
            Then, $Z^{\frac{1}{2}}$ can be built from the unitary
            blocks $Z^{\frac{1}{2}}_j=K_j S_j^{\frac{1}{2}} K_j^{\H}$
            and is itself unitary.
            Here, $S_j^{\frac{1}{2}}$ can be computed as an element-wise square-root
            since $S_j$ is diagonal. 
            Thus, ultimately, $Z^{\frac{1}{2}}$ is unitary and block-diagonal
            and the block sizes correspond to those of $Z$.
            
            \par Finally, the matrices defined by~\cite{PhysRevA.93.062115,Kopylov2025}
            \begin{subequations}
                \begin{align}
                    U &= \tilde{\UU} D, \label{eq:jsdecomp_U_UUD} \\
                    \Psi &= V^{*} D^{*} \label{eq:jsdecomp_Psi_DPsi}
                \end{align}
            \end{subequations}
            fulfill the condition of the joint Schmidt
            decomposition as written in \Cref{eq:jsdecompBeta,eq:jsdecompEtatilde}.

    \section{ASYMMETRY OF THE TRANSFER FUNCTION \texorpdfstring{$\beta$}{β}}%
            \label[appendix]{sec:beta_asym}

        As was discussed in \Cref{sec:single_crystal}, in a general case, 
        the moduli of the higher-order
        input and output modes of a single crystal are no longer 
        necessarily equal at high gain.
        This asymmetry can be explained as originating from an asymmetry in the modulus
        $\abs{\beta\of{q,q'}}$ of the transfer function $\beta$. 
        Note that as discussed in \Cref{sec:jsdecomp}, for non-degenerate 
        singular values, the mode functions are defined up to a constant
        phase. Therefore,  
        if only the moduli of the modes
        or the relative behavior of their phases over $q$
        are of interest, it is sufficient 
        to consider a single Schmidt decomposition of the transfer function $\beta$ instead of the
        joint Schmidt decomposition.

        \begin{figure*}[ht!]
            \centering
            \subfloat{\label{fig:beta_mod_delta_modulus}%
                \includegraphics[width=0.333\linewidth]{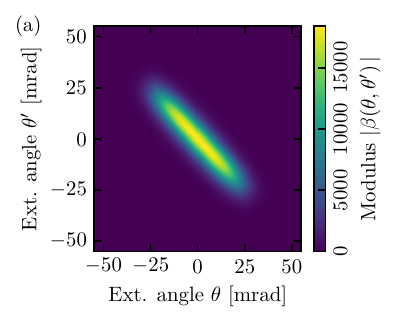}}%
            \hfill%
            \subfloat{\label{fig:beta_mod_delta_moddelta}%
                \includegraphics[width=0.333\linewidth]{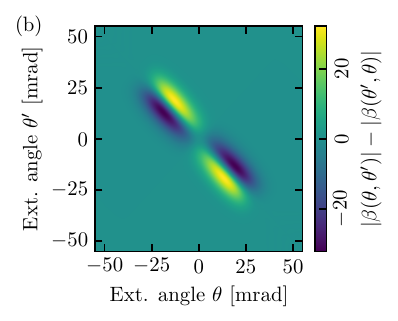}}%
            \hfill%
            \subfloat{\label{fig:beta_mod_delta_phase_phase}%
                \includegraphics[width=0.333\linewidth]{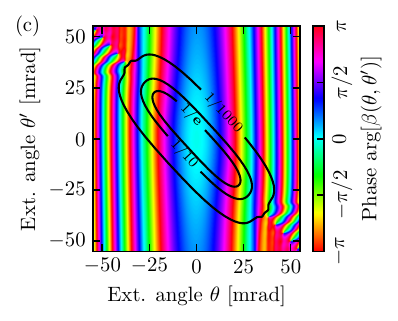}}%
            \caption{\label{fig:beta_mod_delta_phase}%
                (a)~Modulus $\abs{\beta\of{\theta,\theta'}}$ of the transfer
                function $\beta$ 
                at high gain $G_{\mathrm{exp}}=8$.
                (b)~Plot of the expression 
                $\abs{\beta(\theta,\theta')}-\abs{\beta(\theta',\theta)}$,
                which reveals an asymmetry in
                the modulus of the transfer function, whose
                modulus is around of a factor 500 smaller than the peak value of the
                transfer function itself. 
                (c)~Phase $\arg\beta\of{\theta,\theta'}$ of the transfer
                function. The contour lines indicate where the modulus 
                has decayed by $1/\E$, $1/10$ and $1/1000$ compared to
                its maximum value, respectively. As the modulus approaches 
                zero, numerical artifacts in the upper left and lower right
                corner become more visible. However, these values to not
                contribute as much to the function overall.
            }%
        \end{figure*}%

        \par Figure~\ref{fig:beta_mod_delta_phase} shows the modulus and 
        the phase of the transfer function
        $\beta$, as well as the asymmetry in its modulus, which is around
        of a factor $500$ smaller than
        the maximum value of the modulus of the transfer function itself.
        In order to verify that this effect is not a result of a 
        numerical error,
        we have repeated the numerical calculation of the results 
        shown in \Cref{fig:beta_mod_delta_phase}
        using a \emph{Lie group integrator}. 
        All numerical results shown in this work
        were computed using an adaptive-stepsize Runge-Kutta method of order 5(4) as
        implemented in \emph{Scipy}~\cite{2020SciPy-NMeth}.
        However, the regular Runge-Kutta integration methods are
        not explicitly constructed to preserve the symplectic structure of
        the complex-valued $2n\times 2n$ matrix (compare \Cref{sec:jsdecomp})
        \begin{align}
            S\of{L} &= \begin{bmatrix}
                \tilde{\Eta}\of{L} & \Beta\of{L} \\
                \Beta^{*}\of{L} & \tilde{\Eta}^{*}\of{L}
            \end{bmatrix}
        \end{align}
        during the evolution of the transfer functions over $L$.
        Symplectic matrices $S$
        are defined by the fact that fulfill\footnote{In numerical applications, 
        depending on the implementation, $J$ may be replaced by $J/\dd q$, where $\dd q$
        is the step size of the $q$ lattices. This corresponds to the replacement
        of the identity matrix by the Dirac delta distribution in the continuous case.}~\cite{Houde2024}
        \begin{align}\label{eq:sympl_struct_S}
            S^{\T}JS = J,
        \end{align}
        where
        \begin{align}
            J = \begin{bmatrix}
                0 & I_n \\
                -I_n & 0
            \end{bmatrix}
        \end{align}
        is the \emph{canonical symplectic matrix}. 
        If $S$ is a symplectic matrix, its 
        inverse is given by $S^{-1}=J^{-1}S^{\T}J$
        and is itself a symplectic matrix.
        Explicitly evaluating
        the left-hand side of \Cref{eq:sympl_struct_S} for both $S^{-1}$ and $S$ yields the 
        four well-known~\cite{PhysRevA.93.062115,Christ_2013,Kopylov2025,Houde2024} conditions on the
        matrices representing the transfer function:
        \begin{subequations}
            \begin{align}
                \tilde{\Eta}\tilde{\Eta}^{\H} - \Beta\Beta^{\H} &= I_n, \\
                \tilde{\Eta}\Beta^{\T} - \Beta\tilde{\Eta}^{\T} &= 0,
            \end{align}
        \end{subequations}
        and
        \begin{subequations}
            \begin{align}
                \tilde{\Eta}^{\H}\tilde{\Eta} - \Beta^{\T}\Beta^{*} &= I_n, \\
                \tilde{\Eta}^{\H}\Beta - \Beta^{\T}\tilde{\Eta}^{*} &= 0,
            \end{align}
        \end{subequations}
        respectively [compare also \Cref{eq:BBpIeqHH}]. These are the discretized
        versions of Eqs.~(A3a), (A3b), (D3a) and~(D3b) of \refcite{PRR5}, respectively.
        Thus, the symplectic structure of $S$ ensures that the bosonic~\cite{PhysRevA.93.062115} 
        commutation relations of the output plane-wave operators are preserved.

        \par To be more precise, for simplicity, we have used the 
        \emph{Lie-Euler method}~\cite{Celledoni02012022,Iserles_Munthe-Kaas_Nørsett_Zanna_2000},
        which is a variation
        of the Euler method and ensures that the Lie group structure 
        [here, the structure of the symplectic group $\mathrm{Sp}\of{2n,\mathbb{C}}$]
        of the solution of the integro-differential equation is preserved 
        [meaning $S\of{L}\in\mathrm{Sp}\of{2n,\mathbb{C}}$ for all $L$].
        We have found that there are no qualitative differences in the results obtained 
        using the aforementioned Runge-Kutta method (presented in this section and the rest of
        this work) and those obtained using the Lie-Euler method, meaning
        that, in other words, all plots shown in this work would remain unchanged when produced 
        using the Lie-Euler method.

        \par
\begin{figure}[ht!]%
            \centering%
            \includegraphics[width=\linewidth]%
                {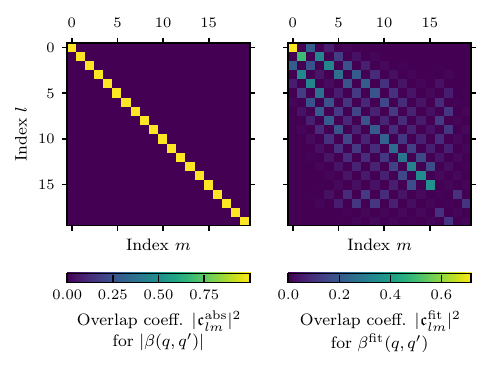}%
            \caption{\label{fig:beta_fit_omat}%
                Left: Modulus squared of the 
                first $20\times 20$ overlap matrix 
                elements $\abs{\occ_{lm}^{\mathrm{abs}}}^2$ for
                the decomposition of $\abs{\beta\of{q,q'}}$ as written in
                \Cref{eq:approx_decomp_takagi_beta}.
                Right: Modulus squared of the 
                first $20\times 20$ overlap matrix 
                elements $\abs{\occ_{lm}^{\mathrm{fit}}}^2$
                for the transfer function $\beta$ composed of its
                modulus and the fit of its phase function as described 
                by \Crefrange{eq:composite_decomp_beta_fit_beta}{eq:composite_decomp_beta_fit_psi}.
            }%
        \end{figure}

 The fact that the moduli of the lower order modes coincide 
        (see \Cref{fig:modesandphasesc1,fig:moremodes})
        can be explained as follows:
        Clearly, \Cref{fig:beta_mod_delta_moddelta,fig:beta_mod_delta_modulus}
        indicate that the modulus of the transfer function $\beta$
        is approximately symmetric: $\abs{\beta\of{q,q'}}\approx\abs{\beta\of{q',q}}$.
        The Schmidt decomposition of the modulus of $\beta$
        can be written as
        \begin{align}\label{eq:approx_decomp_takagi_beta}
            \abs{\beta\of{q,q'}}=\sum_n \sqrt{\Lambda_n^{\mathrm{abs}}}\, u_{n}^{\mathrm{abs}}\of{q} 
                \psi_{n}^{\mathrm{abs}}\of{q'},
        \end{align}
        one can calculate the same-crystal overlap 
        matrix $\occ$ [see \Cref{eq:samecrystaloverlapc}; here, 
        $\occ_{lm}^{\mathrm{abs}}= 
        \int\!\dd q\,u_l^{\mathrm{abs}}\of{q} \left[\psi_m^{\mathrm{abs}}\of{q}\right]^{*}$], 
        see \Cref{fig:beta_fit_omat}
        (left column), where the first $20\times 20$ overlap matrix entries are presented.
        This matrix is almost diagonal, 
        $\abs{\occ_{lm}^{\mathrm{abs}}}^2\approx\delta_{lm}$, which means
        that $\abs{u_{n}^{\mathrm{abs}}\of{q}}\approx\abs{\psi_{n}^{\mathrm{abs}}\of{q'}}$
        and the functions $u_{n}^{\mathrm{abs}}\of{q}$ and $\psi_{n}^{\mathrm{abs}}\of{q'}$ 
        differ only by a constant phase.

        \begin{figure}[ht!]%
            \centering%
            \includegraphics[width=\linewidth]
                {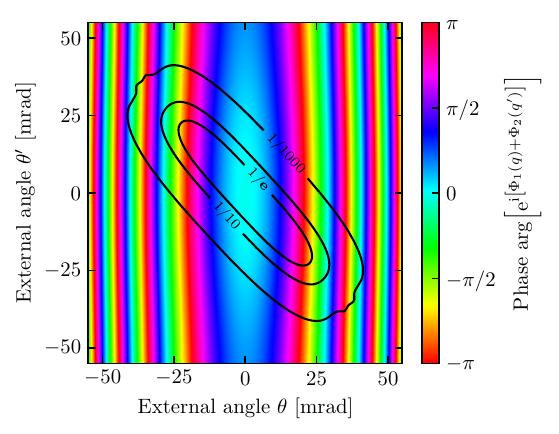}%
            \caption{\label{fig:beta_phase_fit}%
                Plot of the fitted phase distribution 
                $\arg\ofb{\E^{\I\left[\Phi_1\of{q}+\Phi_2\of{q'}\right]}}$,
                where
                $\Phi_j\of{q}=a_{j,1} + a_{j,2} q^2 + a_{j,3} q^4 + a_{j,4} q^6$,
                with $a_{j,k}$ being the fit parameters.
                The fit suggests that $\arg\ofb{\beta\of{q,q'}}$ is approximately factorizable
                in the form $\arg\ofb{\beta\of{q,q'}}=\E^{\I\Phi_1\of{q}}\E^{\I\Phi_2\of{q'}}$.
                The contour lines as shown in \Cref{fig:beta_mod_delta_phase_phase} are
                added as a visual guide regarding the modulus of $\beta$.
                The values of the fitting parameters are
                listed in \Cref{tab:ajk_beta_fit_params}.
            }%
        \end{figure}%

        \begin{table}%
            \renewcommand{\arraystretch}{1.3}%
            \begin{center}%
                \caption{\label{tab:ajk_beta_fit_params}%
                    Fitting parameters for the phase distribution of $\beta$ as shown
                    in \Cref{fig:beta_phase_fit}, in units of
                    $\pi\,\unit{\rad\micro\metre\tothe{2(k-1)}}$.%
                }%
                \begin{tblr}{width=\linewidth,colspec={X[c]X[c]X[c]}}%
                    \hline\hline%
                    $k$ & $a_{1,k}$  & $a_{2,k}$ \\%
                    \hline%
                    1 & -1.000 & 0.995 \\%
                    2 & 31.125 & 1.208 \\%
                    3 & 2.758 & -2.671 \\%
                    4 & -2.119 & 2.100 \\%
                    \hline\hline%
                \end{tblr}%
            \end{center}%
        \end{table}

        \par Furthermore, as illustrated by \Cref{fig:beta_phase_fit},
        the phase distribution $\arg\ofb{\beta\of{q,q'}}$
        is, to a good approximation, factorizable. This means that  
        an approximation to the Schmidt decomposition of $\beta$
        can be written using the fitted phase distribution 
        (see \Cref{fig:beta_phase_fit} for details) as 
        \begin{subequations}
            \begin{align}\label{eq:composite_decomp_beta_fit_beta}
                \begin{split}
                    \beta\of{q,q'} \approx 
                    \beta^{\mathrm{fit}}\of{q,q'} &=
                        \abs{\beta\of{q,q'}}\,\E^{\I\left[\Phi_1\of{q}+\Phi_2\of{q'}\right]} \\
                        &=\sum_n \sqrt{\Lambda_n^{\mathrm{fit}}}\, u^{\mathrm{fit}}_{n}\of{q} 
                            \psi^{\mathrm{fit}}_{n}\of{q'},
                \end{split}
            \end{align}
            where the mode functions are given by
            \begin{align}
                u^{\mathrm{fit}}_{n}\of{q}  &= u_{n}^{\mathrm{abs}}\of{q}\E^{\I\Phi_1\of{q}}, 
                    \label{eq:composite_decomp_beta_fit_u}\\
                \psi^{\mathrm{fit}}_{n}\of{q'} &= \psi_{n}^{\mathrm{abs}}\of{q'}\E^{\I\Phi_2\of{q'}}.
                    \label{eq:composite_decomp_beta_fit_psi}
            \end{align}
            and the eigenvalues are the same as for $\abs{\beta\of{q,q'}}$:
            \begin{align}
                \Lambda_n^{\mathrm{fit}} &= \Lambda_n^{\mathrm{abs}}
            \end{align}
        \end{subequations}
        These mode functions are clearly orthonormal, as required by
        the Schmidt decomposition. Furthermore, they have the same
        modulus as the modes defined by \Cref{eq:approx_decomp_takagi_beta}:
        $\abs{u^{\mathrm{fit}}_{n}\of{q}}=\abs{u_{n}^{\mathrm{abs}}\of{q}}$ and
        $\abs{\psi^{\mathrm{fit}}_{n}\of{q'}}=\abs{\psi_{n}^{\mathrm{abs}}\of{q'}}$.
        As such, of these mode functions $u^{\mathrm{fit}}_{n}\of{q}$ and
        $\psi^{\mathrm{fit}}_{n}\of{q'}$, the lower order ones for the same index
        will have an approximately equal modulus:
        $\abs{u^{\mathrm{fit}}_{n}\of{q}}\approx\abs{\psi^{\mathrm{fit}}_{n}\of{q'}}$.
        However,
        this will not hold for higher
        order modes: Due to asymmetry of $\beta$ and the fact that
        the phase is only approximately separable, the
        matrix $\abs{\occ_{lm}^{\mathrm{abs}}}$ is not fully diagonal,
        which explains the difference in the
        moduli of $u_{15}$ and $\psi_{15}$ as visible in
        \Cref{fig:moremodes}.

        \par Additionally, 
        it immediately follows that the complex behavior of the overlap
        matrix shown in \Cref{fig:modesandphasesc1} for $G_{\mathrm{exp}}=8$
        is partially due to the additional phases $\E^{\I\Phi_1\of{q}}$ and
        $\E^{\I\Phi_2\of{q'}}$ added onto the mode functions.
        It can be shown that the mode functions
        $u^{\mathrm{fit}}_{n}\of{q}$ and $\psi^{\mathrm{fit}}_{n}\of{q'}$
        are very close to the mode functions of the exact decomposition:
        Figure~\ref{fig:beta_fit_omat} (right column) shows the overlap matrix
        coefficients 
        $\occ_{lm}^{\mathrm{fit}} 
        = \int\!\dd q\, u_l^{\mathrm{fit}}\of{q} \left[\psi_m^{\mathrm{fit}}\of{q}\right]^{*}$
        for the modes as constructed 
        in \Cref{eq:composite_decomp_beta_fit_u,eq:composite_decomp_beta_fit_psi}.
        Clearly, the behavior of the overlap matrix is strikingly similar
        to that shown in \Cref{fig:modesandphasesc1}.

    \section{COMPUTATION OF THE VISIBILITY}\label[appendix]{sec:app_vis}
        For fixed~$\delta z$, the full phase dependence of
        the transfer functions of the entire interferometer 
        can be expressed via the transfer functions of the two
        crystals as~\cite{PRR5}:
        \begin{subequations}
            \begin{align}
                \begin{split}
                    \beta^{\left(\mathrm{SU}\right)}\of{q,q'} &= \int\!\dd\bar{q}\,
                        \nophase{\tilde{\eta}}^{\left(2\right)}\of{q,\bar{q}} \beta^{\left(1\right)}\of{\bar{q},q'} \\
                    &\qquad+ \E^{\I\phi} \int\!\dd\bar{q}\,\nophase{\beta}^{\left(2\right)}\of{q,\bar{q}}
                        \left[\tilde{\eta}^{\left(1\right)}\of{\bar{q},q'}\right]^{*},
                \end{split} \\
                \begin{split}
                    \tilde{\eta}^{\left(\mathrm{SU}\right)}\of{q,q'} &= \int\!\dd\bar{q}\,
                        \nophase{\tilde{\eta}}^{\left(2\right)}\of{q,\bar{q}} \tilde{\eta}^{\left(1\right)}\of{\bar{q},q'} \\
                    &\qquad+ \E^{\I\phi} \int\!\dd\bar{q}\,\nophase{\beta}^{\left(2\right)}\of{q,\bar{q}}
                        \left[\beta^{\left(1\right)}\of{\bar{q},q'}\right]^{*},
                \end{split}
            \end{align}
        \end{subequations}
        where we explicitly take the phase out of the transfer functions, considering that the function~$h$ for the integration over
        the second crystal does not contain the phase term~$\E^{\I\phi}$,
        compare \Cref{eq:h2_comp_imperfect} and see the notation
        introduced in \Cref{eq:nophase_notation}.
        With the shorthand
        \begin{subequations}
            \begin{align}
                \nophase{\XX}\of{q,q'} &= \int\!\dd\bar{q}\,
                            \nophase{\tilde{\eta}}^{\left(2\right)}\of{q,\bar{q}} {\beta}^{\left(1\right)}\of{\bar{q},q'}, \\
                \nophase{\YY}\of{q,q'} &= \int\!\dd\bar{q}\,\nophase{\beta}^{\left(2\right)}\of{q,\bar{q}}
                            \left[\tilde{\eta}^{\left(1\right)}\of{\bar{q},q'}\right]^{*},
            \end{align}
        \end{subequations}
        the integral intensity for the interferometer
        can be written as
        \begin{subequations}
            \begin{align}
                \begin{split}
                    \langle\hat{N}_{s,\mathrm{tot}}^{\left(\mathrm{SU}\right)}\rangle &= 
                        \iint\!\dd q\dd q'\,\left(\abs{\nophase{\XX}\of{q,q'}}^2+\abs{\nophase{\YY}\of{q,q'}}^2\right)
                        \\
                        &\qquad+ 2\Re\of{\mathcal{C}\E^{\I\phi}},
                \end{split} \label{eq:nsuto_CCYYCC}
            \end{align}
            where
            \begin{align}
                \mathcal{C} &= \iint\!\dd q\dd q'\,\nophase{\XX}^{*}\of{q,q'}\nophase{\YY}\of{q,q'}.
            \end{align}
        \end{subequations}
        Since, in general,~$\mathcal{C}$ is a complex constant, the positions
        of the bright and dark fringe, 
        where~$\langle\hat{N}_{s,\mathrm{tot}}^{\left(\mathrm{SU}\right)}\rangle$
        is maximized and minimized, respectively, do not, in general, occur at~$\phi=0$
        and~$\phi=\pi$, respectively.
        Instead, the positions are shifted by some phase
        \begin{align}
            \Upsilon &= \arg\of{\mathcal{C}}.
            \label{eq:Upsilon}
        \end{align}
        Usually, this phase offset is relatively small;
        as an example, for the interferometer as
        described in \Cref{sec:unbalanced_interf},
        $\Upsilon\approx -0.0337\pi$.
        The visibility as defined in \Cref{eq:def_vis} can ultimately be 
        written in the form
        \begin{align}\label{eq:form_vis_XY}
            v &= \frac{2\abs{
                \iint\!\dd q\dd q'\,\nophase{\XX}^{*}\of{q,q'}\nophase{\YY}\of{q,q'}
            }}{
                \iint\!\dd q\dd q'\,\left(\abs{\nophase{\XX}\of{q,q'}}^2+\abs{\nophase{\YY}\of{q,q'}}^2\right)
            } \cdot \SI{100}{\percent}. 
        \end{align}
        \par For a perfectly compensated SU(1,1) interferometer
        as discussed in \Cref{sec:perfect_compens_interf},
        we find
        \begin{align}
            \abs{\mathcal{C}} &= 
                \iint\!\dd q\dd q'\,\abs{\nophase{\XX}\of{q,q'}}^2
                = \iint\!\dd q\dd q'\,\abs{\nophase{\YY}\of{q,q'}}^2
                = \AA,
        \end{align}
        where~\cite{PRR5}
        \begin{align}\label{eq:def_AA}
            \AA = \sum_n \Lambda_n^{\left(1\right)} \left(1+\Lambda_n^{\left(1\right)}\right),
        \end{align}
        and therefore, 
        \revA{it follows from \Cref{eq:nsuto_CCYYCC} that}
        $v=\SI{100}{\percent}$, regardless of any
        other parameter. This is also evident from the fact that
        at the dark fringe, which here occurs
        exactly at~$\phi=\pi$, the integral intensity is always
        identically zero, see the discussion in \refcite{PRR5}.

    \section{BEHAVIOR OF THE PHASES OF THE OVERLAP COEFFICIENTS NEAR THE OPTIMAL VISIBILITY}%
            \label[appendix]{sec:gh_delta_z_phase}

        As discussed in \Cref{sec:unbalanced_interf}, for an unbalanced
        SU(1,1) interferometer, the focusing element must be moved some
        distance $\delta z$ away from the optimal position of a balanced 
        SU(1,1) interferometer in order to maximize the interferometric visibility.
        Furthermore, \Cref{fig:omatsplots} shows the behavior of the phases
        of the overlap matrices for the bright and dark fringe for the value of
        $\delta z$ which maximizes the visibility.

        \par
\begin{figure}[t]%
            \centering%
            \includegraphics[width=\linewidth]{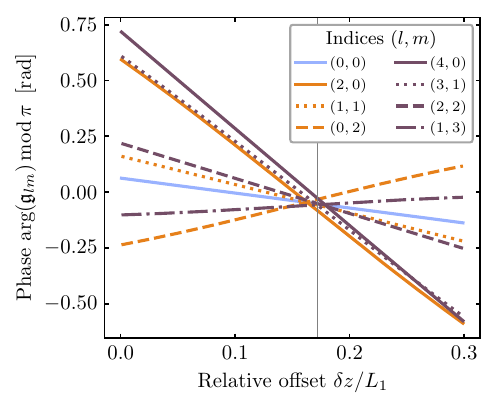}%
            \caption{\label{fig:phase_g_dfbf}Plot of the phase 
            $\arg\of{\ocg_{lm}}$ of the first significant
                entries of the $\ocg$-overlap coefficient matrix for varying offset $\delta z$
                of the focusing element inside the unbalanced SU(1,1) interferometer.
                The vertical thin gray line
                indicates the offset at which the visibility is
                maximized, compare \Cref{fig:visoverdeltaz}.
                As discussed in \Cref{sec:uniq_sdecomp}, the modes 
                and therefore also the $\ocg$-overlap coefficients
                are only defined up to their phase. Therefore, 
                the plotted values are corrected by shifting them
                to the interval $[-\pi/2,\pi/2]$ to achieve a smooth plot. 
                Clearly, at the point of the
                optimal visibility, the phase values of the shown overlap 
                matrix entries almost coincide.
                See also \Cref{fig:phase_h_dfbf}, which discusses the phases
                of the $\och$ overlap matrix 
                elements for the same situation.
                }%
        \end{figure}

 In \Cref{fig:phase_g_dfbf}, this observations is analyzed in more detail 
        in terms of
        the phase of the $\ocg$ overlap matrix for varying 
        $\delta z$. Clearly, as the visibility
        is maximized, the phases of the first few overlap coefficients
        assume approximately the same value. This can also be seen from 
        \Cref{fig:omatsplots}. Since the full dependence on the phase is 
        described by \Cref{eq:g_with_phase_imperfect}, it is clear that
        these phases take approximately the same value for all $\phi$,
        if they assume the same phase for \emph{any} $\phi$.
        This observation
        was also required in \Cref{sec:high_gain_approx_reconstruction}.

        \par
\begin{figure*}%
            \centering%
            \includegraphics[width=\linewidth]{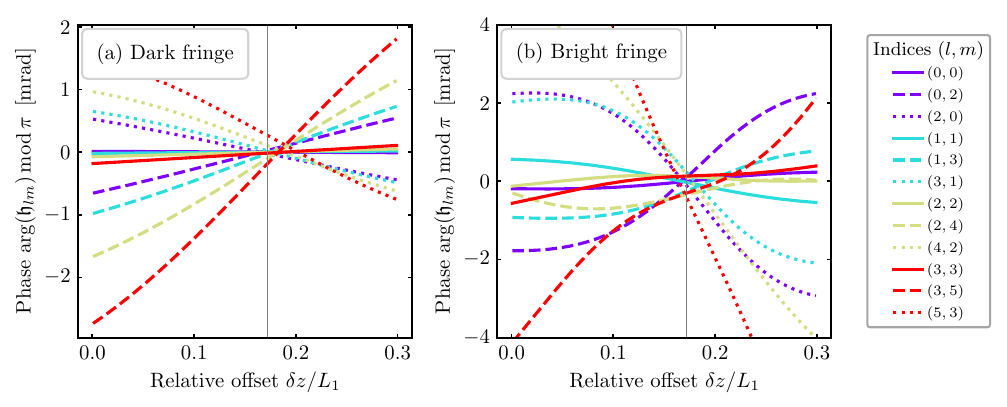}%
            \caption{\label{fig:phase_h_dfbf}Plot of the phase $\arg\of{\och_{lm}}$ of the first significant
                entries of the $\och$-overlap coefficient matrix for varying offset $\delta z$
                of the focusing element inside the unbalanced SU(1,1) interferometer for the (a)~dark fringe
                and (b)~bright fringe. The vertical thin gray line
                indicates the offset at which the visibility is maximized, compare \Cref{fig:visoverdeltaz}.
                As discussed in \Cref{sec:uniq_sdecomp}, the modes and therefore also the $\och$-overlap coefficients
                are only defined up to their phase. Therefore, the plotted values are corrected by shifting them
                to the interval $[-\pi/2,\pi/2]$ to achieve a smooth plot. Clearly, near the point of the
                optimal visibility, the phase values of the shown overlap matrix entries almost coincide.
                Note that unlike  \Cref{fig:phase_g_dfbf}, the y-axis here has its units in $\unit{\milli\rad}$.
                Thus, the phase in the entire shown region is already very close to zero compared to the 
                phase $\arg\of{\ocg_{lm}}$.
                }%
        \end{figure*}

 Similarly, \Cref{fig:phase_h_dfbf} shows the behavior of the phases
        of the $\och$ overlap matrix entries at the bright and dark fringe for varying
        $\delta z$. Compared to the $\ocg$ overlap matrix, the phase values of the 
        $\och$ overlap coefficients are already very close to zero on the entire shown
        region (note that the y-axis here is in \unit{\milli\rad}). Furthermore,
        it is visually clear that for the dark fringe, the point at which
        the phase values are most similar is slightly towards larger
        values of $\delta z$ compared to the point at which the
        visibility is maximized. For the bright fringe, both points coincide.
        
    \section{ACCOUNTING FOR INTERNAL LOSSES}\label[appendix]{sec:lossinclusion}
        It is well known that internal losses which the signal and idler
        fields suffer between the
        two crystals of the SU(1,1) interferometer degrade the interferometer
        performance~\cite{DemkowiczDobrzaski2012,PhysRevA.96.053863}.
        One common way of modeling these losses is by introducing a 
        \emph{virtual}
        beam splitter inside the interferometer whose two input
        ports receive the output fields of the first crystal 
        and the vacuum state~\cite{PhysRevA.96.053863}. This beam splitter is
        virtual in the sense that it only serves as a theoretical
        way of describing the losses and would not be included in
        an experimental implementation of the setup.
        One of the output ports of the beam splitter
        is then sent to the input of the second crystal, while
        the output field of the second output port models the lost radiation
        and is essentially discarded. Note that this description bears similarity
        to the description of lossy PDC provided in \refcite{Kopylov2025}.

        \par Using this description to model the SU(1,1) interferometer
        with losses, it is clear that the transfer functions $\beta^{\left(1\right)}$
        and $\tilde{\eta}^{\left(1\right)}$ describing the first crystal remain
        unchanged, when internal losses are included.
        However, assuming that the beam splitter applies a certain
        amount of losses $\lof$ to the output field of the first crystal,
        the relationship connecting the plane-wave input operators for the
        second crystal to its output operators are, instead of 
        \Cref{eq:as_dagger_sol,eq:ai_dagger_sol}, given by
        \begin{widetext}
            \begin{subequations}
                \begin{align}
                    \begin{split}
                        \hat{a}_s^{\left(2,\mathrm{out}\right)}\of{q_s}
                            &= \sqrt{1-\lof} \left\lbrace\int\!\dd q_s'\,\tilde{\eta}\of{q_s,q_s'}
                            \hat{a}_s^{\left(2,\mathrm{in}\right)}\of{q_s'}
                            +\int\!\dd q_i'\,\beta\of{q_s,q_i'}
                            \left[\hat{a}_i^{\left(2,\mathrm{in}\right)}\of{q_i'}\right]^{\dagger}
                            \right\rbrace \\
                            &\quad+ \sqrt{\lof} \left\lbrace\int\!\dd q_s'\,\tilde{\eta}_0\of{q_s,q_s'}
                            \hat{b}_s^{\left(2,\mathrm{in}\right)}\of{q_s'}
                            +\int\!\dd q_i'\,\beta_0\of{q_s,q_i'}
                            \left[\hat{b}_i^{\left(2,\mathrm{in}\right)}\of{q_i'}\right]^{\dagger}
                            \right\rbrace,
                        \label{eq:as_dagger_sol_lossy}
                    \end{split} \\
                    \begin{split}
                        \left[\hat{a}_i^{\left(2,\mathrm{out}\right)}\of{q_i}\right]^{\dagger} 
                            &= \sqrt{1-\lof} \left\lbrace \int\!\dd q_i'\,
                            \tilde{\eta}^{*}\of{q_i,q_i'}
                            \left[\hat{a}_i^{\left(2,\mathrm{in}\right)}\of{q_i'}\right]^{\dagger}
                            + \int\!\dd q_s'\,\beta^{*}\of{q_i,q_s'}
                            \hat{a}_s^{\left(2,\mathrm{in}\right)}\of{q_s'}
                            \right\rbrace \\
                            &\quad+ \sqrt{\lof} \left\lbrace \int\!\dd q_i'\,
                            \tilde{\eta}_0^{*}\of{q_i,q_i'}
                            \left[\hat{b}_i^{\left(2,\mathrm{in}\right)}\of{q_i'}\right]^{\dagger}
                            + \int\!\dd q_s'\,\beta_0^{*}\of{q_i,q_s'}
                            \hat{b}_s^{\left(2,\mathrm{in}\right)}\of{q_s'}
                            \right\rbrace,
                            \label{eq:ai_dagger_sol_lossy}
                    \end{split}
                \end{align}
            \end{subequations}
        \end{widetext}
        where the newly appearing $\hat{b}_{s/i}$-operators 
        originate from the vacuum input port of the
        beam splitter. These new operators unconditionally commute with
        the $\hat{a}_{s/i}$ operators:
        \begin{subequations}
            \begin{align}
                \comm{\hat{b}_{s/i}\of{q_{s/i}}}{\hat{a}_{s_i}\of{q_{s/i}'}}=0, \\
                \comm{\hat{b}_{s/i}^{\dagger}\of{q_{s/i}}}{\hat{a}_{s_i}\of{q_{s/i}'}}=0, \\
                \comm{\hat{b}_{s/i}^{\dagger}\of{q_{s/i}}}{\hat{a}_{s_i}^{\dagger}\of{q_{s/i}'}}=0, 
            \end{align}
        \end{subequations}
        which is to say that the state belonging to these
        operators is orthogonal to that belonging to the 
        $\hat{a}^{\left(2,\mathrm{in}\right)}_{s/i}$
        operators~\cite{DemkowiczDobrzaski2012}.
        While the
        $\hat{a}^{\left(2,\mathrm{in}\right)}_{s/i}$-operators connect
        the second crystal to the first crystal,
        the $\hat{b}^{\left(2,\mathrm{in}\right)}_{s/i}$-operators
        introduce a new vacuum component to the input of
        the second crystal. Importantly, these operators are
        therefore not connected to the output operators
        via the same transfer functions as the 
        $\hat{a}^{\left(2,\mathrm{in}\right)}_{s/i}$-operators,
        but rather by the newly introduced transfer functions
        which have the subscript~$_0$ added.
        
        \par Plugging \Cref{eq:as_dagger_sol_lossy,eq:ai_dagger_sol_lossy}
        into the operator-value integro-differential equations
        as derived in \refcite{PRR2} (compare also \refcite{PRR5}),
        now yields two sets of coupled integro-differential equation,
        in one of which $\tilde{\eta}$ and $\beta$, and in the other
        of which $\tilde{\eta}_0$ and $\beta_0$
        are each coupled via the transverse wave-vectors
        $q_s$ and $q_i$. The form of each set is the same as written in
        \Cref{eq:integ_diffeqs_eb_beta,eq:integ_diffeqs_eb_eta}.
        Additionally, the both sets require different phase
        matching functions $h$: For the newly introduced 
        $\hat{b}_{s/i}$-operators, the phase matching function
        is that of a single (first) 
        crystal [\Cref{eq:pm_func_c1_ideal,eq:pm_func_c1_nonideal}] 
        instead of that of
        the second crystal of an SU(1,1) interferometer.

        \par As a consequence, it is not longer possible to diagonalize
        the input-output relation in the same way as it was done 
        with the help of the Schmidt operators 
        in \Cref{eq:bogoliubov_plain}. This can be seen from the
        fact that the Schmidt modes resulting from a decomposition
        of $\tilde{\eta}$ and $\beta$ and of $\tilde{\eta}_0$ and $\beta_0$
        will necessarily be different and it is no longer possible to bring
        the plane-wave input-output relations to the form of the 
        Bogoliubov transformations.
        A similar observation was also made in \refcite{PhysRevA.90.023823}
        where input-output relations of a similar structure were obtained.
        Ultimately, this means that the mathematical formalisms of
        \Cref{sec:interferometers,sec:mmsqm} are no longer directly
        applicable in the case of non-vanishing internal losses.

        \par An alternative method, which allows for the partial inclusion
        of the internal losses, is as follows: For the construction
        of the overlap matrices, internal losses are neglected,
        which allows for the utilization of the
        formalisms as developed
        in \Cref{sec:interferometers,sec:mmsqm}.
        For the eigenvalues 
        $\Lambda^{\left(2\right)}_n$ and $\Lambda^{\left(\mathrm{SU}\right)}_n$ 
        however, internal losses can be taken into
        account by computing them via the covariance function. 
        Similar to what was
        described in \refcite{PRR5}, the covariance function for the output
        photons
        \begin{align}\label{eq:def_cov}
            \cov\of{q,q'} = \langle \hat{N}(q)\hat{N}(q') \rangle - \langle \hat{N}(q)\rangle \langle \hat{N}(q') \rangle
        \end{align}
        can be obtained from the output plane-wave operators and
        expressed in terms of $\tilde{\eta}$, $\beta$, $\tilde{\eta}_0$
        and $\beta_0$. If the signal and idler photons are treated
        as distinguishable in some degree of freedom, for example
        their polarization or when they are measured at slightly different
        frequencies but in such a way that the corresponding refractive 
        indices can still be treated as being equal, the eigenvalues
        can be obtained by decomposing the square root of the covariance
        function $\cov\of{q,q'}$, see \refcite{multimodesqueezing,Chekhova2025} for 
        more details. These eigenvalues may then be used instead of the
        eigenvalues obtained from the Schmidt decomposition since they
        take into account the internal losses. It should be noted that this
        approximation is only valid for high gain and requires 
        $\Lambda_n\gg 1$. See also Eq.~(20) of \refcite{PRR2}.

\end{document}